\documentclass[12pt]{article}
\usepackage{amsmath}
\usepackage{amsmath,bm}
\usepackage{url}
\usepackage{tikz}
\usetikzlibrary{shapes.geometric}
\usetikzlibrary{arrows}
\usepackage{amsmath}
\usepackage{amssymb}
\usepackage{hyperref}
\usepackage{rotating}
\usepackage{physics}
\usepackage{epsfig}
\usepackage{graphicx}
\usepackage{color}
\usepackage{cite}
\usepackage{subfig}
\usepackage{hhline}
\usepackage[font=small,labelfont=bf]{caption}

\newcommand{\beq}{\begin{equation}}
\newcommand{\eeq}{\end{equation}}
\newcommand{\ga}{\lower.7ex\hbox{$\;\stackrel{\textstyle>}{\sim}\;$}}
\newcommand{\la}{\lower.7ex\hbox{$\;\stackrel{\textstyle<}{\sim}\;$}}

\hypersetup{
    colorlinks = true,
    citecolor = {blue},
    linkcolor = {blue},
    urlcolor = {blue},
}

\numberwithin{equation}{section}

\begin{document}

\def\jcap{\ref@jnl{J. Cosmology Astropart. Phys.}}

\begin{flushright}
{\tt CPHT-RR020.052023, UMN-TH-4212/23, FTPI-MINN-23/06} \\
\end{flushright}

\vspace{0.2cm}
\begin{center}
{\bf {\large Testing the Scalar Weak Gravity Conjecture in No-scale Supergravity }}

\end{center}
\vspace{0.1cm}

\begin{center}{
{\small\bf Emilian Dudas}$^{a}$,
{\small\bf Tony Gherghetta}$^{b}$,
{\small\bf Keith~A.~Olive}$^{b,c}$, and
{\small\bf Sarunas~Verner}$^{d}$}
\end{center}

\begin{center}
{\small\em $^a$Centre de Physique Th{\'e}orique, {\'E}cole Polytechnique, CNRS and IP Paris, 91128 Palaiseau Cedex, France}\\
{\small\em $^b$School of
 Physics and Astronomy, University of Minnesota, Minneapolis, MN 55455,
 USA} \\
{\small\em $^c$William I. Fine Theoretical Physics Institute, School of
 Physics and Astronomy, \\University of Minnesota, Minneapolis, MN 55455,
 USA} \\
 {\small\em $^d$Institute for Fundamental Theory, Physics Department, University of Florida, Gainesville, FL 32611, USA}

 \end{center}

\vspace{0.1cm}
\centerline{\bf ABSTRACT}
\vspace{0.1cm}

{\small We explore possible extensions of the Weak Gravity Conjecture (WGC) to scalar field theories. 
To avoid 
charged black hole remnants, 
the WGC requires the existence of a particle with a mass $m < g q M_P$, with charge $q$ and U(1) gauge coupling $g$, allowing the decay to shed the black hole charge. Although there is no obvious problem that arises in the absence of a U(1) charge, it has been postulated that gravity must remain the weakest force even when extended to scalar interactions. 
Quantifying this conjecture may be done by comparing scalar and gravitational amplitudes, or as we advocate here by comparing scattering cross sections. In theories with non-trivial field space geometries, by working out examples with perturbation theory around arbitrary field values and performing tadpole resummations, we argue that the conjecture must be applied only at extrema of the scalar potential (when expressed in locally canonical coordinates). 
We consider several toy models in the context of no-scale supergravity and also consider examples of inflationary models. 
 } 

\vspace{0.2in}

\begin{flushleft}
{May} 2023
\end{flushleft}
\medskip
\noindent

\newpage

\section{Introduction}

It has become apparent in recent years that not all effective field theories can be consistently embedded into a theory of quantum gravity. In particular, arguments based on black hole physics and string theory give rise to nontrivial constraints that restrict the observable parameter space of low-energy quantum field theories. Theories which violate these constraints are deemed inconsistent when considered part of the larger, fundamental theory of quantum of gravity and are relegated to what has become known as the {\it swampland}\cite{Vafa:2005ui,Ooguri:2006in}. Various swampland conjectures have been formulated (see \cite{Palti:2019pca} for a review) and used to test the available parameter space of low-energy quantum field theories.

A nontrivial conjecture about a theory of quantum gravity is the weak gravity conjecture (WGC)~\cite{ArkaniHamed:2006dz}, which states that in a U(1) gauge theory with gauge coupling $g$ coupled to gravity, there must exist a charged particle with charge $q$ and mass $m$ satisfying $m\leq g q M_P$ (where $M_P$ is the (reduced) Planck mass). This bound is also consistent with the absence of global symmetries in quantum gravity and the requirement of a maximum charge-to-mass ratio $q/m$, so that extremal mass black holes can decay and leave no charged remnants. Interestingly, the WGC can also be reformulated as the absence of a stable weakly-coupled gravitational bound state~\cite{Palti:2017elp}. In other words, requiring that the Coulomb repulsion is larger than the gravitational attraction for the WGC particle with the largest charge-to-mass ratio is equivalent to the WGC. This is sometimes also referred to as the Repulsive Force Conjecture~\cite{Heidenreich:2019zkl}.

The bound state argument provides one way to generalize the WGC to include scalar fields. For massless (moduli) fields that have cubic interactions with the WGC charged particle, this gives rise to a Coulomb-like attractive force that modifies the WGC condition to be\cite{Palti:2017elp}  
\beq
m^2 + g^{ij} \mu_i \mu_j M_P^2 \le g^2 q^2 M_P^2 \, ,
\label{wgc}
\eeq
where $m \mu_i = \nabla_{t^i} m^2/2$ is the cubic coupling
of the WGC particle to moduli $t^i$ and $\nabla_{t^i}$ is the covariant derivative.
An interesting consequence of \eqref{wgc} occurs when the scalar cubic interaction is tuned to the U(1) gauge coupling.
In this case the mass, $m$, must be hierarchically smaller than the Planck scale suggesting a possible form of UV/IR mixing~\cite{Lust:2017wrl}. However, these generalizations require the presence of a charged particle and are not applicable to purely scalar theories.
Strictly speaking, Eq.~(\ref{wgc}) would imply that stable, uncharged massive particles are forbidden. 
In the limit that the charge vanishes, the conjecture can be salvaged by including repulsive scalar interactions and rewriting Eq.~(\ref{wgc}) as \cite{Palti:2017elp}
\beq
m^2 + g^{ij} \mu_i \mu_j M_P^2  \le \frac12 g^{ij} \nabla_{t^i} \nabla_{t^j} m^2 M_P^2 \,.
\label{wgcr}
\eeq 

Nevertheless, a conjecture solely for massless, scalar fields, partially-motivated in the $g\rightarrow 0$ limit can be formulated by requiring that the magnitude of the scalar force, whether attractive or repulsive, is stronger than gravity~\cite{Palti:2017elp}. The Scalar Weak Gravity Conjecture (SWGC) can be expressed as \footnote{Repulsive interactions can also be included in (\ref{swgc}).}
\beq
m^2  \le g^{ij} \mu_i \mu_j M_P^2 \, ,
\label{swgc}
\eeq
where $m$ is now the mass of a state that has a cubic coupling to the moduli $t^i$.
Unlike the WGC, there are no charged black hole remnant arguments, and therefore this conjecture remains on a weaker footing since it is not clear what goes fundamentally wrong if the scalar force is weaker than gravity. Moreover, this conjecture based on long-range forces only applies to massless scalar fields.

Instead, another form of the conjecture for scalar fields was put forth in Ref.~\cite{Gonzalo:2019gjp} based entirely on elevating the statement of gravity as the weakest force to a general principle. The conjecture applies to any scalar field and gives rise to a condition on the scalar potential of a canonically-normalized scalar field that is motivated from four-point scattering amplitudes. The so-called strong SWGC (sSWGC) is a condition that combines a contact four-point scalar interaction, which is repulsive and proportional to the Dirac delta function, with the long range attractive interactions arising from gravity and the scalar field exchange due to cubic interactions. As written, the condition on the scalar potential was expressed as
\beq
m^2 \frac{\partial^2}{\partial \phi^2} \left(\frac{1}{m^2}\right) M_P^2 \ge 1
\eeq
or 
\beq
2 V'''^2 - V'''' V'' \ge \frac{V''^2}{M_P^2} \, ,
\label{sswgc}
\eeq
where $V' = \partial V/\partial \phi$ and $m^2 = V''$.
This implies one cannot factor out a universal distance dependence and therefore the sSWGC encapsulates a UV/IR mixing effect. In fact, in the case of one real scalar field, the condition is saturated for a potential that can be interpreted as containing both Kaluza-Klein and winding states~\cite{Gonzalo:2019gjp}. Furthermore, it was noticed in~\cite{Gonzalo:2019gjp}, that even simple potentials such as $V(\phi) =  \frac12 m^2 \phi^2+\frac14 \lambda \phi^4$, violate the conjecture near the minimum, $\phi = 0$. This point was further emphasized in \cite{Freivogel:2019mtr} where it was noted that the same potential could be used to describe a dilute gas of non-relativistic atoms and hence would violate this version of the conjecture. For other applications of this and related conjectures see \cite{Scalisi:2018eaz,Kusenko:2019kcu,Shirai:2019tgr,Andriot:2020lea,Gonzalo:2020kke}.

\begin{figure}[ht!]
    \centering
    \includegraphics[width=1\columnwidth]{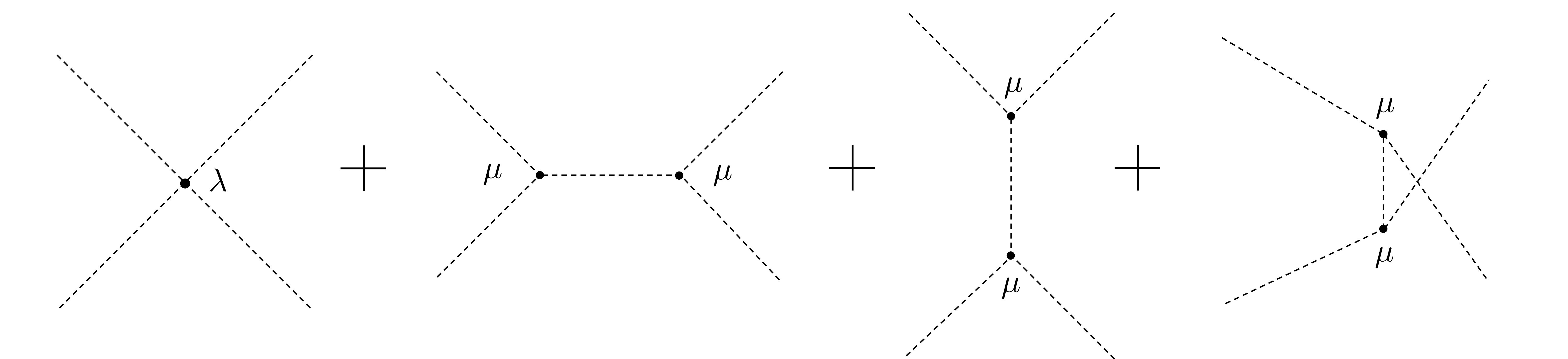}
    \caption{Tree level diagrams contributing to the four-point amplitude.}
    \label{simpletree}
\end{figure}

The sSWGC of Ref.~\cite{Gonzalo:2019gjp} was also scrutinized in Ref.~\cite{Benakli:2020pkm}, where it was suggested to consider the non-relativistic limit of four-point scattering amplitudes and to compare the strength of the scalar and gravitational forces at the same energy scale in an attempt to make sense of the UV/IR condition. In 
particular, this involved comparing the short-range scalar contact interactions with the $s$-channel contact interaction of gravity at arbitrary scalar field values. This requires computing the scattering amplitudes for the scalar field fluctuations. In terms of the potential this modifies Eq.~(\ref{sswgc}) to read as
\beq
\left|\lambda - \frac53  \mu^2 \right| \ge \frac32 \frac{m^2}{M_P^2} \, ,
\label{bsswgc}
\eeq
where $\lambda$ corresponds to the four-point contact interaction, the factor of $\frac53$ accounts for $s, t,$ and $u$-channel diagrams in the non-relativistic limit and the factor of $\frac32$ corresponds to single 
graviton exchange in the $s$-channel. These factors are easily understood from the tree level diagrams shown in Fig.~\ref{simpletree}. In the nonrelativistic limit, the scalar exchange diagrams contribute ($\frac13,-1,-1$) for the $(s,t,u)$-channel diagrams.  The absolute value in \eqref{bsswgc} accounts for either net repulsive or attractive forces, so long as they are stronger than gravity. However, if Eq.~(\ref{bsswgc}) is applied 
at arbitrary field values, fluctuations about these values induce generic linear terms in the scalar potential. Away from the potential extrema, these linear terms can contribute to the potential scattering, but were not included in the examples considered in Ref.~\cite{Benakli:2020pkm}. 

A further reformulation of the sSWGC was presented in Ref.~\cite{Gonzalo:2020kke} where it was proposed that the pair production rate into charged particles of mass $m$ by any force mediator (spin 0 or 1) should be larger than the equivalent gravitational production rate. Inspired by the pair production rate by massless moduli, Ref~\cite{Gonzalo:2020kke} also modified the condition \eqref{sswgc} to
\beq
\left|\xi V'''^2 - V'''' V''\right| \ge \frac{V''^2}{M_P^2} \, ,
\label{eq:newsswgc}
\eeq
where absolute values are inserted and $\xi$ is an order one constant. The pair production argument, which involves an interaction rate rather than an amplitude, gives $\xi=1$ in contrast to $\xi=2$ from \eqref{sswgc}, whereas the amplitude for the nonrelativistic scattering of a single scalar field gives $\xi=\frac{5}{3}$ as argued from \eqref{bsswgc}. The approach taken in Ref.~\cite{Gonzalo:2020kke} is that $\xi$ is not particularly relevant and \eqref{eq:newsswgc} is thought of as a swampland condition in the sense that it becomes trivial when $M_P\rightarrow \infty$. However, the issue of whether the simple recipe $m^2\rightarrow V''$ is consistent or how to  generalize the condition to multiple scalar fields remain unanswered questions.

In this paper, we further investigate the sSWGC and generalize the interpretation of the constraint in terms of scattering amplitudes and interaction rates (or cross sections).
To compute the four-point scattering amplitudes we employ the novel techniques developed in Ref.~\cite{dgov}. In general, we may have several scalar fields with nontrivial kinetic terms which require a canonical normalization in order to compute scattering amplitudes \cite{Alonso:2015fsp,Cohen:2021ucp,Alonso:2021rac,Cheung:2021yog,Cheung:2022vnd,Helset:2022tlf}. As shown in Ref.~\cite{dgov}, we use Riemann normal coordinates (RNC) to canonically normalize the kinetic terms. This allows us to calculate scattering amplitudes and test the scalar conjectures for the various examples in no-scale supergravity \cite{no-scale}. Because of the K\"ahler geometry inherent in supergravity theories, the use of K\"ahler normal coordinates (KNC) was also advocated in \cite{dgov}. Indeed in theories with complex fields the use of KNC can greatly simplify the computation of amplitudes (when supersymmetry is unbroken). These techniques are reviewed in Section \ref{sec:review}.

Furthermore, as discussed in Ref.~\cite{dgov}, under RNC (or KNC) field transformations, the scattering amplitudes are only invariant at the potential extrema where the linear terms vanish. When the linear terms are present, tadpole contributions to the four-point scattering must be included, and we argue that the four-point scattering at an arbitrary field value is then equivalent to computing the scattering amplitude at one of the potential extrema. This means that the four-point scattering, and hence the scalar WGC, only needs to be checked at the scalar potential extrema. In Section \ref{sec:tadpoles}, we examine in some detail the role of tadpoles. We first show that in simple models involving one or two real scalar fields, invariance of the scattering amplitudes under field redefinitions is lost   
when tadpoles are ignored. Of course at potential extrema where tadpole contributions are absent, invariance is maintained. We also consider a perturbative example, which allows us to include tadpole contributions consistently in our expansion parameter, and show that at next-to-leading order the amplitudes computed away from the minimum are equal to the amplitude at the minimum when tadpoles are included. 
Because we are formulating the sSWGC in terms of amplitudes, in general, unless we have a way of summing up all tadpole contributions, we will restrict our attention to computing amplitudes at extrema. We adapt an argument from \cite{dnps} using the Newton Tangent method to argue this point. 

In Section \ref{sec:grav}, we consider real (and complex)
scalar fields minimally coupled to gravity through the energy momentum tensor. Assuming a Minkowski background, we compute the $s$-channel gravitational scattering amplitudes which are then compared to scalar interactions. Thus we further restrict our calculation to minima of the potential with (near) vanishing vacuum energy.  In general, these scattering amplitudes will
contain an angular dependence, and hence when strictly comparing amplitudes, the conjecture is only satisfied in particular directions.
To avoid this issue, we also integrate over the scattering angle, effectively calculating a scattering cross section for each example we consider.  The spirit of the sSWGC would then require the total scalar cross section to be larger than the gravitational cross section.
 This captures the sense of the sSWGC reformulation in \cite{Gonzalo:2020kke} arguing that gravitational interaction rates (here given in terms of a cross section) are smaller than all other competing rates. We label this the $\sigma$SWGC.

In Section \ref{sec:simple}, we consider several
examples that contain one or two complex scalar fields. 
The amplitudes for these examples were computed in \cite{dgov}. Since we are necessarily dealing with multiple real scalar fields that have non-trivial kinetic terms, we cannot express the constraint as simply as in Eq.~(\ref{bsswgc}). Instead, we can consider
\beq
\left| \mathcal{A} \right|_{\rm scalar} \ge \left| \mathcal{A} \right|_{\rm gravity} \, ,
\label{compamp}
\eeq
which is effectively the general form for Eq.~(\ref{bsswgc}) for a specific process (i.e., with specific initial and final states).  Then as noted above, we can integrate these amplitudes (squared), and sum over final states to obtain
\beq
\sigma_{\rm scalar} \ge \sigma_{\rm gravity} \, .
\label{compcs}
\eeq
In all of the simple examples considered, some form of Eq.~(\ref{compcs}) is satisfied whereas there are instances where (\ref{compamp}) is not, particularly since there is in general a dependence on the scattering angle in the scattering amplitude. 

In the WGC, it sufficed to have one particle satisfying Eq.~(\ref{wgc}). 
In the scalar version, one could similarly argue that it is sufficient to have at least one initial state and final state interaction which is stronger than gravity. As will see below this condition ($\sigma_{11}$SWGC) is satisfied in all of the examples we consider and is not very constraining. A stronger version of this would require the existence of a scalar field for which the total inclusive cross section is greater than the gravitational cross section ($\sigma_{1i}$SWGC). This too is satisfied in all of our examples and $\sigma_{1i}$SWGC implies $\sigma_{11}$SWGC. Still stronger versions of the conjecture might require the interactions of all scalars to be stronger than gravity in at least one final state channel ($\sigma_{a1}$SWGC) or summed over all final states  ($\sigma_{ai}$SWGC).
The latter is the most restrictive and $\sigma_{ai}$SWGC implies $\sigma_{a1}$SWGC. In the examples considered, we will comment on which of these conjectures is satisfied.

In Section \ref{sec:infl}, we consider two
inflationary models motivated from no-scale supergravity. Both examples exhibit Starobinsky-like inflation \cite{Staro}, but originate from different superpotentials and as such have different mass spectra. 
At the minimum of the potential,
both models satisfy the $\sigma_{ai}$SWGC, (i.e. Eq.~(\ref{compcs}) for inclusive cross sections for all initial states). However, both models exhibit a limiting case extremum which is a saddle point as the inflaton takes infinite field values. 
In one of the two examples, the $\sigma_{ai}$SWGC is violated at the saddle point, for a limited range in center of mass energies (though $\sigma_{1i}$SWGC is not),  whereas in the other, the $\sigma_{ai}$SWGC is always satisfied. 
Our conclusions are summarized in Section \ref{sec:concl}.

\section{Review of Scattering Amplitudes in No-scale Supergravity}
\label{sec:review}

As shown in \cite{dgov}, the calculation of scattering amplitudes in $\mathcal{N}=1$ supergravity theories can be greatly simplified by making a field redefinition to K\"ahler normal coordinates. Here, we briefly outline that procedure. 

Given a K\"ahler potential, $K(Z^I,\bar{Z}^{\bar I})$, and a superpotential, $W$, we can write the following Lagrangian for the scalar sector of the theory as \cite{Nilles:1983ge}
\begin{equation}
    \label{lagcomplex}
    \mathcal{L} \; = \; K_{I \bar{J}}(Z, \bar{Z}) \partial_{\mu} Z^I \partial^{\mu} \bar{Z}^{\bar{J}} - V(Z, \bar{Z}) \, ,
\end{equation}
where $Z^I$ is a complex scalar field and $K_{I \bar{J}} \equiv \; \partial_{i} \partial_{\,\bar{j}} K$.  The scalar potential can be expressed as
\begin{equation}
    \label{eq:effpotsugra}
    V(Z, \bar{Z}) \; = \; e^G \left[G_I G_{\bar{J}} K^{I \bar{J}}  - 3 \right] \; = \; K_{I \bar{J}} F^I F^{\bar{J}} - 3e^G \, , 
\end{equation}
where
\begin{equation}
    G \equiv K + \ln|W|^2 \, 
\end{equation}
and 
\begin{equation}
    F^I \; = \; - e^{G/2} G^{I \bar{J}} G_{\bar{J}} \, .
\end{equation}
The inverse K\"ahler metric is expressed with upper indices and $K^{I \bar{J}}K_{M \bar{J}} = \delta^I_M$. For a general K\"ahler metric, the calculation of scattering amplitudes can be quite cumbersome as the fields $Z^I$ may not have canonical kinetic terms, thus complicating the expressions for masses and propagators, etc. 

As in the general theory of relativity, given a non-trivial space-time metric, it is always possible to choose a local set of coordinates such that the metric is Minkowskian. Of course, the necessary coordinate transformation is space-time dependent. These coordinates are often referred to as Riemann Normal Coordinates (RNC). 
For a theory with real scalar fields and a non-trivial background field metric, we can also transform to RNC by a suitable field redefinition defined at any point in field space \cite{Higashijima:2000wz,Higashijima:2002fq, Hatzinikitas:2000xe,dgov}. 
Here, we will restrict our attention to theories described by a K\"ahler geometry and the appropriate field redefinition involves K\"ahler Normal Coordinates (KNC). 

The transformation to KNC can be applied at any point in field space,
and we take $Z^I = w^I + z^I$,
where $z^I$ is the fluctuation about a background 
field value, $w^I$. 
As shown in Refs.~\cite{Higashijima:2000wz,Higashijima:2002fq}, one can introduce a holomorphic field redefinition of the field fluctuations $z^I$
\begin{align}   
     \label{eq:transkah1}
     z^I \longrightarrow z^{\alpha} - \sum_{N=2}^{\infty} \frac{1}{N!} \Gamma^{\alpha}_{~\beta_1 \beta_2 \ldots \beta_N} \Big|_w z^{\beta_1}z^{\beta_2} \ldots z^{\beta_N} \, ,
\end{align}
where
\begin{equation}
    \Gamma^{\alpha}_{~\beta_1 \beta_2 \ldots \beta_N}  \; = \; \nabla_{\beta_1} \nabla_{\beta_2} \ldots \nabla_{\beta_{N-2}} \Gamma^{\alpha}_{~\beta_{N-1} \beta_N}\,,
\end{equation}
with the covariant derivatives only acting on the lower indices. The capital Latin letters $I,J, \ldots$ denote the original basis and the Greek letters $\alpha, \beta, \ldots$ denote the K\"ahler normal coordinate basis. 
In (\ref{eq:transkah1}), the Christoffel symbols evaluated at the background field value should be regarded as coefficients of the non-linear transformation. 

The transformation to K\"ahler normal coordinates diagonalizes the K\"ahler metric and canonically normalizes the complex fields. The additional transformation $z^{\alpha} \rightarrow K^{-1/2}_{z^{\alpha} z^{\alpha}} z^{\alpha}$ ensures that $K_{\alpha \bar{\beta}}(w, \bar{w}) = \delta_{\alpha \bar{\beta}}(w, \bar{w})$. Thus, in K\"ahler normal coordinates, the general two-derivative interaction Lagrangian with $N$ massive complex scalar fields~(\ref{lagcomplex}) becomes
\begin{equation}
    \label{eq:KNClagrangian}
    \mathcal{L} \; = \; \left( \delta_{\alpha \bar{\beta}} + R_{\alpha \bar{\beta} \gamma \bar{\delta}} z^{\gamma} \bar{z}^{\bar{\delta}}  + \ldots \right) \partial_{\mu} z^{\alpha} \partial^{\mu} \bar{z}^{\bar{\beta}} - V(z, \bar{z}) \, ,
\end{equation}
where in terms of the K\"ahler potential, the K\"ahler curvature tensor $R_{\alpha \bar{\beta} \gamma \bar{\delta}}=K_{\alpha \bar{\beta},\gamma \bar{\delta}}-K^{\mu\bar{\nu}} K_{\mu\bar{\beta},\bar{\delta}}K_{\alpha\bar{\nu},\gamma}$ involves just ordinary partial derivatives.

The kinetic terms in the Lagrangian (\ref{eq:KNClagrangian}) lead to a
four-point scattering amplitude involving two complex and two complex conjugate fields given by
\begin{equation}
A_{4,\rm kin}^{z_{\alpha_1} z_{\alpha_2} {\bar z}_{\bar \alpha_3} {\bar z}_{\bar \alpha_4} } \; = \; R^{\alpha_1 \bar{\alpha}_3 \alpha_2 \bar{\alpha}_4} s_{12}  \ , 
\label{eq:complex1} 
\end{equation}      
along with similar amplitudes obtained by permuting the indices and crossing symmetry. The scattering amplitudes arising from the potential terms can be easily obtained by 
applying the field redefinition (\ref{eq:transkah1}) in the scalar potential (\ref{eq:effpotsugra}). Then the amplitudes can be determined by taking ordinary partial derivatives with respect to the KNC complex fields.
For example, 
one obtains the following four-point amplitudes involving complex fields
\begin{align}
    A_{4, \rm pot}^{z_{\alpha_1} z_{\alpha_2} \bar{z}_{\bar{\alpha}_3} \bar{z}_{\bar{\alpha}_4}} \; = \;
     & - V^{\alpha_1 \bar{\alpha}_3 \alpha_2 \bar{\alpha}_4} - V^{\bar{\alpha}_3 \alpha_1 \bar{\alpha}_4 \alpha_2} + \frac{1}{2} (V^{\alpha_1 \alpha_2 \bar{\alpha}_3 \bar{\alpha}_4} + V^{\bar{\alpha}_3 \bar{\alpha}_4 \alpha_1 \alpha_2}) \nonumber \\
    -\sum_\beta & \left( \frac{V^{\alpha_1 \alpha_2 \rho_\beta}V_{\rho_\beta}{}^{\bar{\alpha}_3 \bar{\alpha}_4}}{s_{12}-m_{\beta\bar{\beta}}^2}+ \frac{V^{\alpha_2 \bar{\alpha}_3 \rho_\beta}V_{\rho_\beta}{}^{\alpha_1 \bar{\alpha}_4}}{s_{14}-m_{\beta\bar{\beta}}^2}+\frac{V^{\alpha_1 \bar{\alpha}_3 \rho_\beta}V_{\rho_\beta}{}^{\alpha_2 \bar{\alpha}_4}}{s_{13}-m_{\beta \bar{\beta}}^2} \right) \, ,
\end{align}
where the indices are complex $\rho_{\beta} = \beta, {\bar \beta}$. Note that the expressions for the contractions over $\rho_\beta$ are shorthand for example,  $V^{\alpha_1 \alpha_2 \rho_\beta}V_{\rho_\beta}{}^{\alpha_3 \alpha_4} = V^{\alpha_1 \alpha_2 \bar\beta}V_{\beta}{}^{\alpha_3 \alpha_4} + V^{\alpha_1 \alpha_2 \beta}V_{\bar\beta}{}^{\alpha_3 \alpha_4}$. Recall also that in this basis, the mass matrix is diagonal, with $V_{\alpha \beta} = V_{\bar{\alpha} \bar{\beta}} = 0$, and that indices in this basis can be raised and lowered with a Kronecker delta. For more detail, see Ref.~\cite{dgov}. 

It is sometimes more convenient to express the Lagrangian in terms of real scalar fields. 
If we use the field expansion $z^{\alpha} = \frac{1}{\sqrt{2}} \left(\chi^{\alpha} + i \xi^{\alpha} \right)$ and $\bar{z}^{\bar{\alpha}} = \frac{1}{\sqrt{2}} \left(\chi^{\bar{\alpha}} - i \xi^{\bar{\alpha}} \right) = \bar{z}^{\alpha} = \frac{1}{\sqrt{2}} \left(\chi^{\alpha} - i \xi^{\alpha} \right)$, the kinetic terms of the Lagrangian become
\begin{eqnarray}
    \mathcal{K}_{\rm{KNC}} & = & \frac{1}{2}\partial_{\mu} \chi^{\alpha} \partial^{\mu}\chi_{\alpha} + \frac{1}{2}\partial_{\mu} \xi^{\alpha} \partial^{\mu}\xi_{\alpha} \nonumber\\
    &&+ \frac{1}{4}R_{\alpha \bar{\beta} \gamma \bar{\delta}} \left[ ( \partial_{\mu} \chi^{\alpha} \partial^{\mu} \chi^{\bar{\beta}}  + \partial_{\mu} \xi^{\alpha} \partial^{\mu} \xi^{\bar{\beta}}) (\chi^{\gamma} \chi^{\bar{\delta}} +  \xi^{\gamma} \xi^{\bar{\delta}} -i (\chi^{\gamma} \xi^{\bar{\delta}} -  \xi^{\gamma} \chi^{\bar{\delta}}))\right. \nonumber \\
    && \left. - ( \partial_{\mu} \chi^{\alpha} \partial^{\mu} \xi^{\bar{\beta}}  
    - \partial_{\mu} \xi^{\alpha} \partial^{\mu} \chi^{\bar{\beta}}) (\chi^{\gamma} \xi^{\bar{\delta}} -  \xi^{\gamma} \chi^{\bar{\delta}} +i (\chi^{\gamma} \chi^{\bar{\delta}} +  \xi^{\gamma} \xi^{\bar{\delta}})) \right]~.
    \end{eqnarray}
Using the above kinetic terms of the Lagrangian, we can write the four-point scattering amplitudes. For example, 
\begin{eqnarray}
    A_{4, \, \rm kin}^{\chi_{\alpha_1}\chi_{\alpha_2}\chi_{\alpha_3}\chi_{\alpha_4}}&=&  A_{4, \, \rm kin}^{\xi_{\alpha_1}\xi_{\alpha_2}\xi_{\alpha_3}\xi_{\alpha_4}} = \frac{1}{4}\left(R^{\alpha_1 \bar{\alpha}_3 \alpha_2 \bar{\alpha}_4} s_{12} + R^{\alpha_1 \bar{\alpha}_2 \alpha_3 \bar{\alpha}_4} s_{13} + R^{\alpha_1 \bar{\alpha}_2 \alpha_4 \bar{\alpha}_3} s_{14}\right) + \rm{c.c.}  \nonumber \\
     \label{Amp4kin:KNC}
\end{eqnarray}
Other combinations of the real fields $\chi$ and $\xi$ as incoming and outgoing states are possible and given in \cite{dgov}. The potential contributions in this basis are analytically cumbersome (these are also given in full in \cite{dgov}) but again are easily obtained from the field redefinition and splitting the complex fields to their real components. 

The full scalar amplitude is given by
$\mathcal{A}_{\rm scalar} =  \mathcal{A}_{4, \, \rm kin} + \mathcal{A}_{4, \, \rm pot}$
for a specified set of initial states $\{\alpha_1,\alpha_2\}$ and final states $\{\alpha_3, \alpha_4\}$. In the reformulated sSWGC, we then require $|\mathcal{A}|_{\rm scalar}>|\mathcal{A}|_{\rm gravity}$ which is discussed in detail in Section \ref{sec:grav} below.

\section{The role of tadpoles}
\label{sec:tadpoles}

\subsection{Simple scattering amplitudes and non-linear field redefinitions}
\label{sec:simptad}

It was shown in Ref.\cite{dgov} that scattering amplitudes in no-scale supergravity do not remain invariant with respect to field redefinitions (in general non-linear) whenever tadpole terms are present. That is, whenever amplitudes are evaluated away from a local extremum. 
 Here we show that the same problem exists in very simple toy models involving one or two real scalar fields.

 Consider first a single, massive real scalar field with quartic self-interactions described by the Lagrangian
 \beq
 {\cal L} \; = \;\frac{1}{2} \partial_\mu \phi \partial^\mu \phi - \frac{1}{2} m_\phi^2 \phi^2 -\frac{1}{4!}\lambda_\phi \phi^4 \, .
 \label{L1f}
 \eeq
 The four-point scattering amplitude for fluctuations about the potential minimum, $\phi = 0$
 is immediately seen to be simply
 \beq
{\cal A}_{4\phi}^{(0)} = \lambda_\phi \, .
\label{A1-4}
 \eeq
 Next, we perform a non-linear field redefinition $\phi \to \phi + a \phi^3$ where $a$ is a real parameter. 
 The Lagrangian up to quartic interactions becomes
 \beq
 {\cal L} \; = \;\frac{1}{2} \partial_\mu \phi \partial^\mu \phi - \frac{1}{2} m_\phi^2 \phi^2 + 3 a \phi^2 \partial_\mu \phi \partial^\mu \phi  - \frac{1}{4!} \left(24 a m_\phi^2 +\lambda_\phi\right) \phi^4 \, .
 \label{L1f2}
 \eeq
 Note that the field redefinition was chosen with a linear term in $\phi$ so that at the quadratic level the Lagrangians \eqref{L1f} and \eqref{L1f2} are identical.
 The four-point amplitude derived from the potential terms is now
 \beq
{\cal A}_{4\phi}^{\rm pot} = 24 a m_\phi^2 + \lambda_\phi \, .
 \eeq
However, kinetic terms of the form $f \phi^2 \partial_\mu \phi \partial^\mu \phi$ also 
provide a contribution, $-8 f m_\phi^2$, so that with $f = 3 a$,
\beq
{\cal A}_{4\phi}^{\rm kin} = -24 a m_\phi^2 \,,
\eeq
and the total amplitude is again given by Eq.~(\ref{A1-4}). Note that adding a cubic term to the Lagrangian of the form $(\mu/3!) \phi^3$
would not affect the invariance for fluctuations about the zero background value.
The cubic interaction would add $\phi$ exchange diagrams (as in Fig.~\ref{simpletree}) to the four-point amplitude which becomes ${\cal A}_{4\phi} = \lambda_\phi - \frac53 \mu^2/m_\phi^2$ in the non-relativistic limit. A simple calculation shows that the same result is obtained after the non-linear transformation.

Perhaps a more interesting example involves two real scalar fields, $\phi$ and $\eta$, where we assume a similar Lagrangian for $\eta$ 
with mass $m_\eta$ and self-interaction $\lambda_\eta$. Since the fields are decoupled, there is no non-zero amplitude for 
$\phi \phi \to \eta \eta$ scattering (or any permutation of these fields).
Performing the transformations 
$\phi \to \phi + a \phi^3 + b \phi \eta^2$ and $\eta \to \eta + c \eta^3 + d\eta \phi^2$, where $a,b,c,d$ are constants, 
the Lagrangian, up to quartic interactions, becomes
\begin{eqnarray}
{\mathcal L} & = & \frac{1}{2} \partial_\mu \phi \partial^\mu \phi + (3 a \phi^2 +b \eta^2) \partial_\mu \phi \partial^\mu \phi + 2(b+d) \eta \phi \partial_\mu \phi \partial^\mu \eta  \nonumber \\
& & + \frac{1}{2} \partial_\mu \eta \partial^\mu \eta + (d \phi^2 + 3 c \eta^2) \partial_\mu \eta \partial^\mu \eta \nonumber \\
& & - \frac12 m_\phi^2 (\phi^2 + 2 a \phi^4 + 2 b \phi^2 \eta^2)  -\frac{1}{4!}\lambda_\phi \phi^4 \nonumber \\
& & - \frac12 m_\eta^2 (\eta^2 + 2 c \eta^4 + 2 d \phi^2 \eta^2) - \frac{1}{4!}\lambda_\eta \eta^4 \, .
\end{eqnarray}
The amplitude for $\phi \phi \to \phi\phi$ is, as before, ${\cal A}_{4\phi} = \lambda_\phi$, and similarly ${\cal A}_{4\eta} = \lambda_\eta$.
For $\phi \phi \to \eta \eta$,  the contributions from the potential terms are
\beq
{\cal A}_{2\phi2\eta}^{\rm pot} = 4 (b m_\phi^2 +  d m_\eta^2) \, ,
 \eeq
and from the kinetic terms
\beq
{\cal A}_{2\phi2\eta}^{\rm kin} = 2 b (s - 2m_\phi^2) + 2 d (s - 2m_\eta^2) - (2b + 2d) s =-4 (b m_\phi^2 +  d m_\eta^2)\, ,
\eeq
where $s = (p_1 + p_2)^2$ is the Mandelstam variable. 
The sum of these contributions ${\cal A}_{2\phi2\eta}^{\rm pot}+{\cal A}_{2\phi2\eta}^{\rm kin} =0$.
Of course, this should not come as a surprise.
We are calculating a scattering amplitude in a well-defined background corresponding to the potential minimum at $\phi = \eta = 0$ and we expect the invariance of the scattering amplitudes with respect to field redefinitions because the original and field-transformed Lagrangians are identical at the quadratic level.

Let us now shift the field in Eq.~(\ref{L1f}) away from its minimum by taking $\phi =v+\varphi$.
While the kinetic terms for the shifted field are unchanged (i.e. they remain minimal), the potential contains additional constant, linear, quadratic and cubic terms. The nonzero linear terms generated at generic field values $v\neq0$, imply that the scattering of the fluctuations $\varphi$, is no longer occurring at a potential minimum. Clearly, the background field value is now time dependent and the process is not strictly equivalent to potential scattering around a stable vacuum. Nevertheless, in a Fourier expansion of the time-dependent field, we can ignore the time-dependent modes of the scalar field and just consider the time-independent zero mode to define a scattering process. This approximation is only valid where $\dot v$ can be neglected compared to the interaction time scale, but will be sufficient for our purposes.

With these caveats, the scattering amplitude for $v \ne 0$ in the non-relativistic limit becomes
\beq
{\cal A}_{4 \varphi}^{(v)} = \lambda_\phi - \frac53 \frac{\lambda_\phi^2 v^2}{m_\phi^2 + \frac12 \lambda_\phi v^2} \, ,
\label{Av}
\eeq
where the second term includes contributions from $s$, $t$, and $u$ exchange diagrams generated by the cubic coupling, $\lambda_\phi v$ with a mass squared, $m_\phi^2 + \frac12 \lambda_\phi v^2$, and we have ignored any contributions from the linear (tadpole) terms. 
If we now perform the non-linear redefinition on the fluctuation, 
$\varphi \to \varphi + a \varphi^3$, the kinetic and potential contributions to the four-point contact term conspire to still give $\lambda_\phi$, however, the $\varphi$-exchange terms in the non-relativistic limit lead to a total amplitude
\beq
{\cal A}_{4\varphi}^{(v)} = \lambda_\phi - \frac53 \frac{v^2(\lambda_\phi+6a ( m_\phi^2 + \frac{1}{6}\lambda_\phi v^2))^2}{m_\phi^2 + \frac12 \lambda_\phi v^2} \, ,
\eeq
which disagrees with Eq.~(\ref{Av}) for all $a \ne 0$, when $v \ne 0$. 
We can only conclude that the inclusion of the tadpole terms is necessary for the invariance of the amplitudes. 

\begin{figure}[ht!]
    \centering
    \includegraphics[width=1\columnwidth]{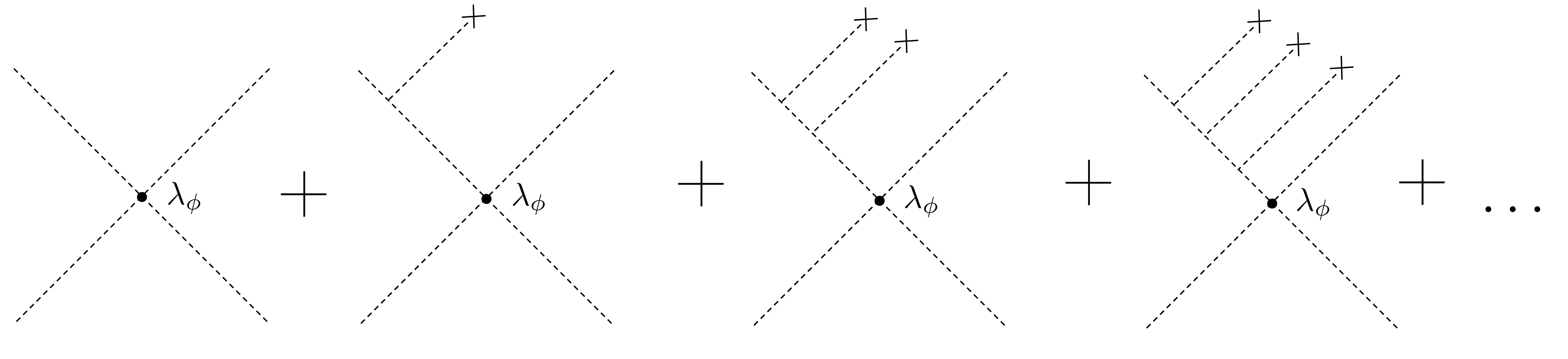}
    \caption{Corrections to the four-point contact diagram including an infinite series of tadpole diagrams from cubic vertices. }
    \label{tadp1}
\end{figure}

\begin{figure}[ht!]
    \centering
    \includegraphics[width=1\columnwidth]{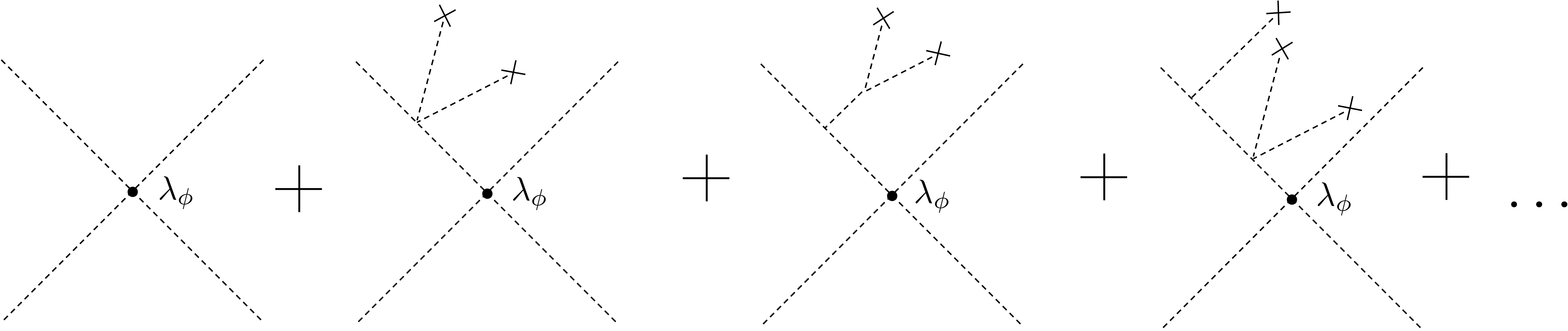}
    \caption{Additional correction diagrams involving both cubic and quartic vertices.  }
    \label{tadp2}
\end{figure}

The latter statement is much easier expressed than calculated.
Consider for example, the above single field model with $m_\phi  = 0$. 
When we shift the field to a non-zero vacuum value, we generate a constant, a tadpole term, a mass term, and a cubic term.
Imagine including a single tadpole contribution to the four-point contact term, as shown in the second diagram in Fig.~\ref{tadp1}.
The contribution to the amplitude is
\beq
\Delta {\cal A}_{4\varphi}^{(1ti)} = \lambda_\phi \frac{(\lambda_\phi v) (\frac16 \lambda_\phi v^3)}{\frac12 \lambda_\phi v^2 (p_i^2-\frac12 \lambda_\phi v^2)}  =  \frac{\frac13\lambda_\phi^2 v^2}{ p_i^2-\frac12 \lambda_\phi v^2}  \, ,
\eeq
where
$\lambda_\phi v$ corresponds to the cubic vertex connecting to the tadpole, $\frac16 \lambda_\phi v^3$ is the linear term and 
the denominator accounts for the two propagators with mass $\frac12 \lambda_\phi v^2$ and momentum $p_i$ of the external leg containing the tadpole. Since this is a non-perturbative contribution, we
must also include diagrams with an arbitrary number of tadpoles generated either through the contact or cubic couplings. Examples of the series of diagrams needed to be summed are shown in Figs.~\ref{tadp1} and \ref{tadp2}.  In addition, 
we would have to include all of the contributions of the tadpoles to the $s$, $t$, and $u$ exchange diagrams generated by the shifted field. We have not found a way to sum all of these diagrams in complete generality.
We provide however a specific example, where the contribution of the tadpoles can be included perturbatively. 

\subsection{A perturbative toy example including tadpoles} 
\label{sec:perttad}

Let us start again with the Lagrangian in Eq.~(\ref{L1f}) and shift the field to an nonzero vacuum value with 
$ \epsilon \equiv v/m_\phi \ll 1$. That is, we consider only background field values close to the potential minimum at $\phi = 0$. The Lagrangian (\ref{L1f}) becomes
\begin{eqnarray}
 {\cal L} &= &\frac{1}{2} \partial_\mu \varphi \partial^\mu \varphi -\frac{1}{2}\epsilon^2 m_\phi^4 \left(1+\epsilon^2\frac{\lambda_\phi}{12}\right) -m_\phi^3 \epsilon \left(1+\epsilon^2\frac{\lambda_\phi}{6}\right)\varphi ~\nonumber\\&&- \frac{1}{2} m_\phi^2 \left(1+\epsilon^2\frac{\lambda_\phi}{2}\right)\varphi^2 -\frac{1}{3!}\epsilon \lambda_\phi m_\phi\varphi^3
 -\frac{1}{4!}\lambda_\phi \varphi^4 \, .
 \label{L1eps}
 \end{eqnarray}
Consider the four-point scattering amplitude.
To order $\epsilon^0$ (i.e., ignoring the tadpoles and exchange diagrams), there is only a contact term contributing
\beq
{\cal A}_{4\varphi}^{(0)} = \lambda_\phi~,
\label{A4}
\eeq
to the total amplitude as in Eq.~(\ref{A1-4}). For the next correction to the four-point amplitude,
at order $\epsilon^2$, there are $s$, $t$, and $u$ exchange diagrams which contribute
\beq
{\cal A}_{4\varphi}^{(3)} = \epsilon^2 \lambda_\phi^2 m_\phi^2 \left(\frac{1}{s-m_\phi^2} +\frac{1}{t-m_\phi^2}+\frac{1}{u-m_\phi^2} \right) = -\frac53 \epsilon^2  \lambda_\phi^2~,
\eeq
where the second equality corresponds to the non-relativistic limit ($s\approx 4 m_\phi^2, t\approx 0, u\approx 0$).
These arise from the same diagrams shown in Fig.~\ref{simpletree} with $\mu = \epsilon \lambda_\phi m_\phi$.

\begin{figure}[ht!]
    \centering
    \includegraphics[width=1\columnwidth]{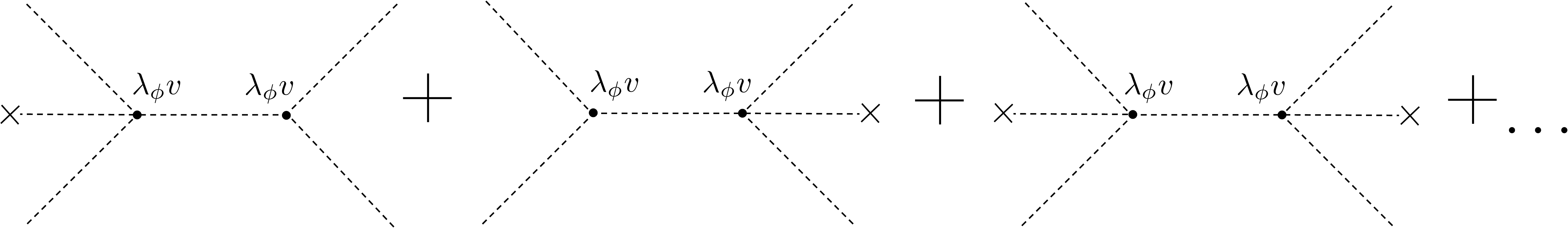}
    \caption{Tadpole contributions to $\mathcal{A}_{4\varphi}^{(31t)}$ and $\mathcal{A}_{4\varphi}^{(32t)}$ in Eqs.~(\ref{31t}) and (\ref{32t}).}
    \label{tadpe}
\end{figure}

Next, it is possible to place a tadpole on one or both of the vertices of the exchange diagrams. For one tadpole insertion,
\beq
{\cal A}_{4\varphi}^{(31t)} = -2 \epsilon^2\lambda_\phi^2  m_\phi^2 \left(\frac{1}{s-m_\phi^2} +\frac{1}{t-m_\phi^2}+\frac{1}{u-m_\phi^2} \right) = \frac{10}{3}  \epsilon^2 \lambda_\phi^2 ~,
\label{31t}
\eeq
where the factor of two in front accounts for inserting the tadpole on either vertex and the second equality is the nonrelativistic limit.
Inserting tadpoles on each vertex gives
\beq
{\cal A}_{4\varphi}^{(32t)} = \epsilon^2 \lambda_\phi^2 m_\phi^2 \left(\frac{1}{s-m_\phi^2} +\frac{1}{t-m_\phi^2}+\frac{1}{u-m_\phi^2} \right) = -\frac53 \epsilon^2 \lambda_\phi^2 \,,
\label{32t}
\eeq
where again the nonrelativistic limit has been taken in the second equality. The three diagrams contributing to Eqs.~(\ref{31t}) and (\ref{32t}) are shown in Fig.~\ref{tadpe}. 
One readily sees that ${\cal A}_{4\varphi}^{(3)} + {\cal A}_{4\varphi}^{(31t)} + {\cal A}_{4\varphi}^{(32t)} = 0$.

This cancellation at order $\epsilon^2$ in the four-point amplitude is already apparent in the cubic vertex at order $\epsilon$. Indeed, the effective cubic vertex obtained from inserting the tadpole in the contact term is
\beq
    {\cal A}_{3\varphi}^{(1t)} = -\lambda_\phi m_\phi \epsilon \left(1-\frac{\lambda_\phi}{2} \epsilon^2+\dots\right)~.
\eeq
Thus, the effective cubic vertex ${\cal A}_{3\varphi}^{(0)}+{\cal A}_{3\varphi}^{(1t)} = {\cal O}(\epsilon^3)$ where from the Lagrangian \eqref{L1eps} ${\cal A}_{3\varphi}^{(0)}=\lambda_\phi m_\phi \epsilon$.

Furthermore, there are contributions to order $\epsilon^2$ from adding tadpoles to the contact interaction.
There is 1) a contact term 
with a single tadpole
\beq
{\cal A}_{4\varphi}^{(41t)} = - \lambda_\phi \frac{(m_\phi^2 v)( \lambda_\phi v)}{m_\phi^2 (p_i^2-m_\phi^2)} = -\epsilon^2\frac{\lambda_\phi^2 m_\phi^2}{p_i^2-m_\phi^2}~,
\eeq
as in the second diagram in Fig.~\ref{tadp1}.
In this expression, $p_i$ refers to the momentum of the leg where the tadpole is placed, which can be any one of the incoming or outgoing momenta. A pair of tadpoles can also be joined to the contact coupling on any of the four legs as in the second diagram of Fig.~\ref{tadp2}, giving
\beq
{\cal A}_{4\varphi}^{(42t)} = \lambda_\phi \frac{\lambda_\phi (m_\phi^2 v)^2}{m_\phi^4 (p_i^2-m_\phi^2)} = \epsilon^2 \frac{\lambda_\phi^2 m_\phi^2}{p_i^2-m_\phi^2}~,
\eeq
so that their sum ${\cal A}_{4\varphi}^{(41t)} + {\cal A}_{4\varphi}^{(42t)} = 0$. 

Thus to order $\epsilon^2$, 
all tadpole terms are cancelled
and the total amplitude at background field values away from (but close to) the minimum is exactly equal to the amplitude at the minimum, ${\cal A}_{\rm tot} = {\cal A}_{4\varphi}^{(0)}$. 
All other diagrams with tadpoles are of higher order in $\epsilon$. 
We have also checked that this cancellation is maintained to order $\epsilon^4$.
This result, if true to all orders in $\epsilon$ and for arbitrarily large values of $\epsilon$ (i.e., non-perturbative),
would imply that scattering amplitudes calculated at arbitrary field values 
are simply equivalent to the amplitudes calculated at the minimum. If true, this is an important result as we have already seen, the 
calculation of amplitudes at arbitrary field values when ignoring the tadpole contributions leads to results which are dependent on particular field redefinitions 
which renders the notion of an invariant scattering amplitude meaningless. 

\subsection{Tadpole resummations and the Newton tangent method: Feynman versus Newton}

Given that a general scalar potential can have several extrema, an important question is the following. If a scattering amplitude is computed at an arbitrary scalar field value, to which extremum will the amplitude converge to, after tadpole resummations?
The answer was provided in \cite{dnps}, where 
it was shown that tadpole resummations for scalar VEVs have a remarkably simple interpretation using 
Newton's tangent method, a highly convergent iterative procedure to find roots of non-linear algebraic equations. In the single field case, it amounts to considering the function $V'(\Phi)$, whose zeroes are the extrema of the scalar potential. One starts at a scalar field value $\Phi^{(0)}$ with a “wrong vacuum”, and proceeds with a sequence of iterations
determined by the zeroes of the curves tangent to the derivative of the potential
\beq
V'(\Phi)= V' (\Phi^{(n)}) +V^{''} (\Phi^{(n)})
 \ ( \Phi -  \Phi^{(n)} )  \ . \label{eq:ntm1}
\eeq
The subsequent field values are defined recursively from $V'(\Phi=\Phi^{(n+1)})=0$ to give 
\beq 
 \Phi^{(n+1)} =  \Phi^{(n)} -
\frac{V' (\Phi^{(n)})}{V^{''} (\Phi^{(n)})}
\ , \label{eq:ntm2}
\eeq 
with $\Phi^{(n)}$ denoting the $n$-th iteration of the initial field value $\Phi^{(0)}$. 

\begin{figure}[ht!]
    \centering
    \includegraphics[width=0.8\columnwidth]{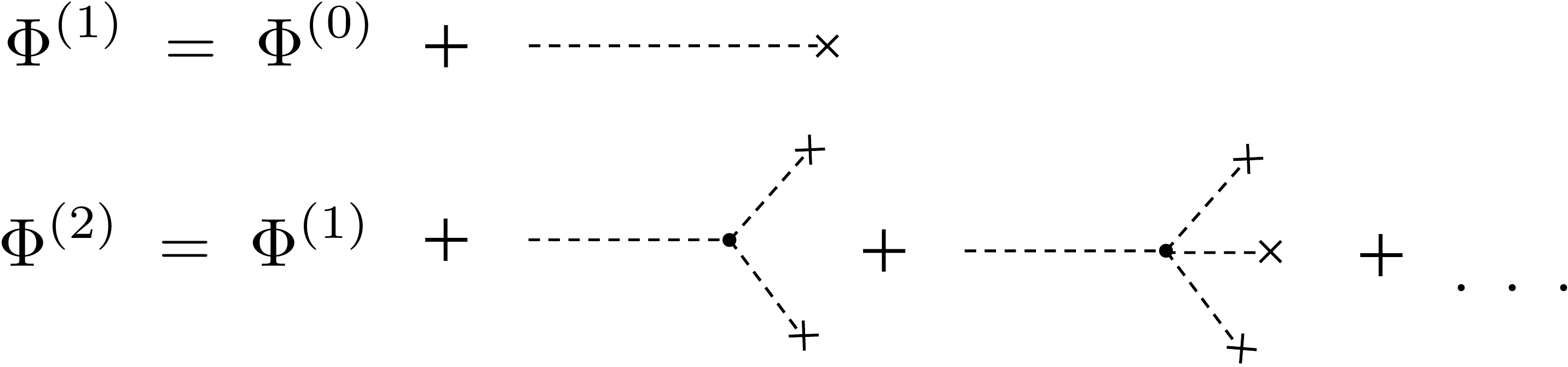}
    \caption{The Feynman diagrams corresponding to the first few terms in the Newton tangent method.}
    \label{fig:newtmethod}
\end{figure}

By denoting $V' (\Phi^{(0)})\equiv V'$, $V^{''} (\Phi^{(0)}) \equiv V^{''}$ and similarly for the higher derivatives,  the first two iterations of the field value are given explicitly by 
\begin{eqnarray} 
&& \Phi^{(1)} = \Phi^{(0)} - \frac{V'} {V^{''}} \ , \nonumber \\
&& \Phi^{(2)} =  \Phi^{(1)} - \frac{ V' (\Phi^{(1)})} {V^{''} (\Phi^{(1)})}   = 
\Phi^{(0)} - \frac{V'} {V^{''}} \left[  
1 + \frac{\frac{(V') V^{'''} }{2 (V^{''})^2} -  \frac{(V')^2 V^{''''}}{6 (V^{''})^3} }{1 - \frac{V' V^{'''} }{(V^{''})^2} + 
\frac{(V')^2 V^{''''}}{2 (V^{''})^3} } \right] \ \nonumber\,, \\
&& \qquad= 
\Phi^{(0)} - \frac{V'} {V^{''}} - 
\frac{(V')^2 V^{'''} }{2 (V^{''})^3} + 
\frac{(V')^3 V^{''''}}{6 (V^{''})^4} + \cdots 
\ , \label{eq:ntm3}
\end{eqnarray}  
where in the last line an expansion is made in the number of tadpole insertions. It can be shown that each term in the tadpole expansion (\ref{eq:ntm3}) corresponds to a specific Feynman diagram, including its correct combinatorial factor. The Feynman diagrams for the terms in \eqref{eq:ntm3} are shown in Figure~\ref{fig:newtmethod}. It is clear from  (\ref{eq:ntm3}) that the tangent method implements an infinite resummation of diagrams (analogous to Dyson resummation in perturbation theory), with better convergence properties towards an extremum. In general, as is well known for the Newton tangent method (NTM) and exemplified in \cite{dnps} for some scalar potentials, the iterated field value always ends at an extremum. If the initial field value is sufficiently close to an extremum, then the iteration always converges to this extremum. However, for field values far from an extremum, the limit of the iteration depends on the form of the scalar potential (in particular, derivatives, extrema and inflection points), and one can end in a far away extremum. The number of iterations is a priori infinite, but for special tuned initial field values, the iteration can end after one or two steps, since an iteration can ``shoot" directly onto an extremum!   

The upshot of the method is that starting from a given initial field value, one can graphically find the extremum one is driven to upon the tadpole diagram resummations. If one restricts to a specific initial field range, only a subset of all extrema can be reached, whereas if one is interested in all possible initial field values, one can generically end up at any extremum of the scalar potential. 

\begin{figure}[ht!]
    \centering
    \includegraphics[width=0.8\columnwidth]{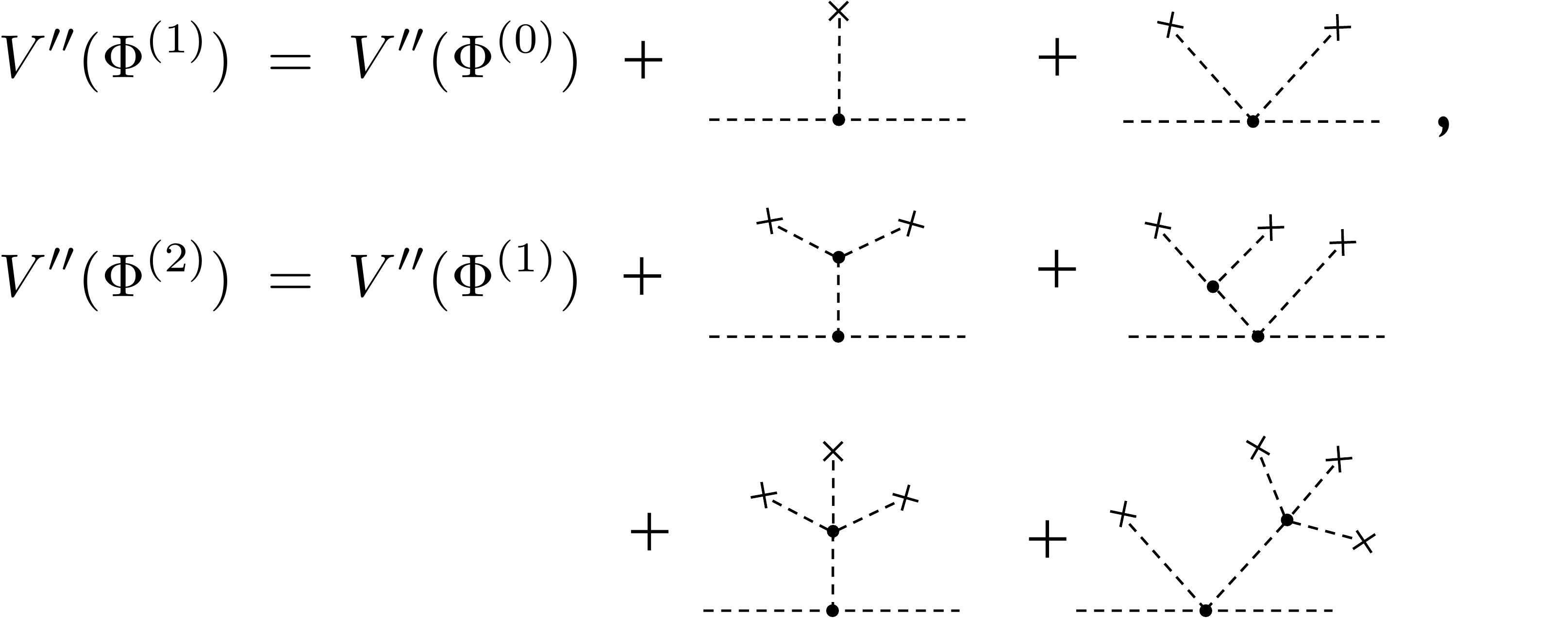}
    \caption{The Feynman diagrams corresponding to the first iteration given by Eq.~(\ref{eq:ntm5}) (top panel) and the second iteration given by Eq.~(\ref{eq:ntm6}) (bottom panel) using Newton's tangent method.}
    \label{fig:newtmethod2}
\end{figure}

The tangent method can also be applied to higher derivatives of the scalar potential, or more generally in the quantum case to the one-point irreducible (1PI) diagrams. Each iteration of the field value $\Phi^{(n)}$ defines a corresponding iteration for these diagrams and a posteriori for the whole effective potential. Let us consider the second derivative. The NTM iteration in this case is defined by 
\beq
V^{''} (\Phi^{(n+1)}) = 
V^{''} \left(\Phi^{(n)}- \frac{V' (\Phi^{(n)})}{V^{''} (\Phi^{(n)})}\right) \ . \label{eq:ntm4}
\eeq 
The first iteration is given by
\beq
V^{''} (\Phi^{(1)}) = 
V^{''} \left(\Phi^{(0)}- \frac{V' (\Phi^{(0)})}{V^{''} (\Phi^{(0)})}\right) \ = \ V^{''}
- \frac{V' V^{'''}}{V^{''}} +  
\frac{(V')^2 V^{''''}}{2 (V^{''})^2} +\dots \ , \label{eq:ntm5}
\eeq 
corresponding to the diagrams shown in the top panel of Fig.~\ref{fig:newtmethod2}. Working out the numerical factors one finds that the first iteration already gives the {\it exact result} for the second derivative in the correct vacuum, $m^2$. The second iteration starts from
\beq
V^{''} (\Phi^{(2)}) = 
V^{''} \left(\Phi^{(1)}- \frac{V' (\Phi^{(1)})}{V^{''} (\Phi^{(1)})}\right) \ = \ V^{''} (\Phi^{(1)})
- \frac{V'(\Phi^{(1)}) V^{'''} (\Phi^{(1)})}{V^{''}(\Phi^{(1)})} +  
\frac{(V'(\Phi^{(1)}))^2 V^{''''}(\Phi^{(1)})}{2 (V^{''}(\Phi^{(1)}))^2} +\dots \ . \label{eq:ntm6}
\eeq
Using the first iteration (\ref{eq:ntm5}), one finally obtains 
\begin{eqnarray}
V^{''} (\Phi^{(2)}) &=& V^{''} (\Phi^{(1)}) - \frac{V'}{V^{''}} 
\frac{ \frac{V' V^{'''}}{2 (V^{''})^2} - \frac{(V')^2 V^{''''}}{6(V^{''})^3}}
{1-\frac{V' V^{'''}}{(V^{''})^2} + \frac{(V')^2 V^{''''}}{2(V^{''})^3}}
\left( 
V^{'''} - \frac{V' V^{''''}}{ V^{''}} \right) \,,
\nonumber \\
& \simeq & V^{''} (\Phi^{(1)})
- \Bigg[  \frac{(V')^2 V^{'''}}{2 (V^{''})^3} - \frac{(V')^3 V^{''''}}{6(V^{''})^4}   \Bigg] \left( 
V^{'''} - \frac{V' V^{''''}}{ V^{''}} \right) + \cdots \ , \label{eq:ntm7}
\end{eqnarray} 
corresponding to the Feynman diagrams displayed at the bottom panel of Fig.~\ref{fig:newtmethod2}. We find that the contributions of the diagrams from the second iteration cancel precisely among themselves. The next iteration only comes from expanding the denominator (first line of (\ref{eq:ntm7})) in the second iteration, and the corresponding diagrams are obtained graphically with additional tadpole insertions obtained in the second iteration. This leads to a ``non-renormalization" theorem: since the diagrams obtained in the second iteration exactly cancel amongst themselves, the higher order iterations will also cancel, and therefore the first iteration already provides the exact 1PI two-point function at zero-momentum (physical mass) in the correct vacuum. 

\begin{figure}[!ht]
    \centering
    \includegraphics[width=0.8\columnwidth]{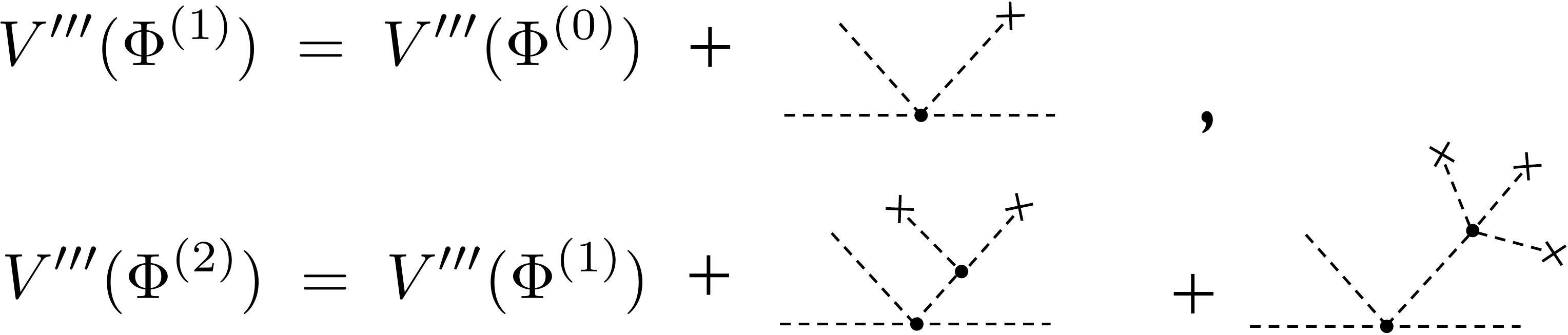}
    \caption{The Feynman diagrams corresponding to the first iteration given by Eq.~(\ref{eq:ntm9}) (top panel) and the second iteration given by Eq.~(\ref{eq:ntm10}) (bottom panel) using Newton's tangent method.}
    \label{fig:newtmethod3}
\end{figure}

It is also instructive to work out the case of the three-point function $V^{'''}$. 
The iteration in this case is
\beq
V^{'''} (\Phi^{(n+1)}) = 
V^{'''} \left(\Phi^{(n)}- \frac{V' (\Phi^{(n)})}{V^{''} (\Phi^{(n)})}\right) =
V^{'''} (\Phi^{(n)}) - \frac{V' (\Phi^{(n)})}{V^{''} (\Phi^{(n)})}  V^{''''} (\Phi^{(n)}) +\dots \ , \label{eq:ntm8}
\eeq 
where $\dots$ refer to higher-order derivatives which vanish for the example in Section~\ref{sec:perttad}.
The first iteration leads to 
\beq 
V^{'''} (\Phi^{(1)}) =
V^{'''} - \frac{V'} {V^{''}}  V^{''''} +\dots  \ , \label{eq:ntm9}
\eeq 
corresponding to the diagram shown in the top panel of Fig.~\ref{fig:newtmethod3}.
The second iteration leads to
\begin{eqnarray}
 V^{'''} (\Phi^{(2)})& =&
V^{'''} \left(\Phi^{(1)} - \frac{V' (\Phi^{(1)})}{V^{''} (\Phi^{(1)})}\right) \,,\nonumber \\ 
& =& V^{'''} (\Phi^{(1)}) 
- \frac{V'}{V^{''}} 
\frac{ \frac{V' V^{'''}}{2 (V^{''})^2} - \frac{(V')^2 V^{''''}}{6(V^{''})^3}}
{1-\frac{V' V^{'''}}{(V^{''})^2} + \frac{(V')^2 V^{''''}}{2(V^{''})^3}}
V^{''''} +\dots\ , \label{eq:ntm10}
\end{eqnarray}
corresponding to the diagrams illustrated in the bottom panel of 
Fig.~\ref{fig:newtmethod3}. 

\begin{figure}[!ht]
    \centering
    \includegraphics[width=0.8\columnwidth]{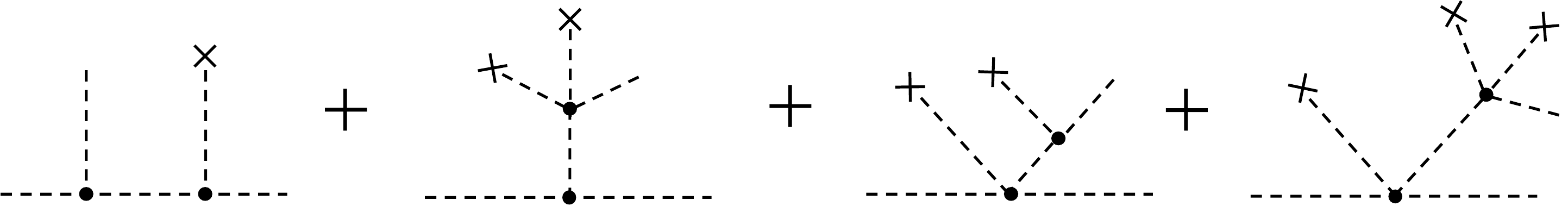}
    \caption{Diagrams that do not contribute to the tadpole resummations in the cubic vertex, since they have a non-zero momentum flowing in the internal propagator. 
    }
    \label{fig:newtmethod4}
\end{figure}

Note that an interesting new feature arises in this case. Whereas in the case of the two-point function the NTM iteration provides all Feynman diagrams with tadpole insertions in perturbation theory, in the case of three-point functions there are four additional Feynman diagrams in Fig. \ref{fig:newtmethod4} that are not present in the NTM iterations shown in Fig.~\ref{fig:newtmethod3}. We find that first of all, the four diagrams in Fig.~\ref{fig:newtmethod4} cancel amongst themselves. Secondly, by adding momenta on the external lines, NTM iterations only generate diagrams where internal propagators have zero momenta, since they are always connected to tadpole insertions. This is indeed the case for the diagrams in Fig.~\ref{fig:newtmethod3} and are therefore the analog of 1PI diagrams of the effective action in perturbation theory. The four different
diagrams in Fig. \ref{fig:newtmethod4} have internal propagators carrying non-zero momentum, since they are not uniquely connected to tadpole insertions and therefore are the analog of reducible 1PI diagrams for the effective action. One can also check that the numerical factors in the NTM iteration do match the two diagrams in  Fig.~\ref{fig:newtmethod3}, whereas some diagrams in  Fig.~\ref{fig:newtmethod4} have different (compared to the NTM iteration) numerical factors.  From this viewpoint, the case of the four-point function ($V^{''''}$)  in the model defined in Eq.~(\ref{L1eps}) is very simple. The NTM iteration in this case is trivial since $V^{'''''}=0$ and therefore all possible Feynman diagrams with tadpole insertions should be reducible. Indeed, by an explicit check, one finds that all diagrams with four-external legs and tadpole insertions have at least one internal propagator with a non-zero momentum, since they are not uniquely connected to tadpole insertions.  Therefore, they do not contribute to the renormalization of the self-coupling, although they should be taken into account in computing physical four-point amplitudes.

Physical amplitudes contain both reducible and irreducible diagrams. Since irreducible diagrams converge by themselves to the result at extrema of the scalar potential, it is clear that all diagrams, reducible and irreducible, will do the same. Whereas we checked this explicitly in Section~\ref{sec:perttad} for the example at hand, we believe that the results of this section
add strong evidence that this is a general result for perturbation theory amplitudes evaluated at an arbitrary field value.

\section{Gravitational scattering amplitudes}
\label{sec:grav}

To test the SWGC, we must compare the amplitudes from non-gravitational interactions
with the amplitude induced by $s$-channel graviton exchange. The gravity $t$ and $u$ channels are not included because they give rise to a long-range Coulomb 
interaction and therefore cannot be compared with the effective short range (Dirac delta function) scalar interaction in the non-relativistic limit. 
Only the gravity $s$-channel gives a short range interaction, which can then be compared with the scalar interaction~\cite{Benakli:2020pkm}.

We start with the Lagrangian (see e.g., \cite{hol}) coupling a real scalar field $\phi$ to gravity 
\beq
{\cal L}= - \frac{1}{M_P}h_{\mu \nu}T^{\mu \nu}_\phi  \, ,
\eeq
where the metric is expanded about Minkowski space-time with $g_{\mu \nu} = \eta_{\mu \nu} + h_{\mu\nu}/2M_P$ and $h_{\mu\nu}$ is the canonically-normalized perturbation. The energy-momentum tensor for a complex scalar field takes the standard form 
\beq
\label{eq:enmomtensor1}
T_{z}^{\mu \nu} =
\left(\partial^\mu z^{\alpha} \partial^\nu \bar{z}_{\bar{\alpha}}+ \partial^\nu z^{\alpha} \partial^\mu \bar{z}_{\bar{\alpha}} \right) -
\eta^{\mu \nu} \left[
\partial^\lambda z^{\alpha} \partial_\lambda \bar{z}_{\bar{\alpha}} -V(z, \bar{z}) \right] + \ldots \, ,
\eeq
where ... refers to couplings of $z^\alpha$ to additional degrees of freedom in the supergravity Lagrangian.
The scattering amplitude for $z^{\alpha_1} \bar{z}^{\bar{\alpha}_1} \to z^{\alpha_2} \bar{z}^{\bar{\alpha}_2}$ then takes the form
\beq
{\cal A}^{z_{\alpha_{1}} \bar{z}_{\bar{\alpha}_1} z_{\alpha_2} \bar{z}_{\bar{\alpha}_2}} = \frac{1}{M_P^2} M_{\mu \nu}^{\,\alpha_2 \bar{\alpha}_2} \, \Pi^{\mu \nu \rho \sigma} \, M_{\rho \sigma}^{\, \alpha_1 \bar{\alpha}_1} \, .
\label{Agravcomplex}
\eeq
Here the graviton propagator, in de Donder gauge, is given by
\begin{equation}
 \Pi^{\mu\nu\rho\sigma}(k) = \frac{1}{2s}\left(\eta^{\rho\nu}\eta^{\sigma\mu} + 
\eta^{\rho\mu}\eta^{\sigma\nu} - \eta^{\rho\sigma}\eta^{\mu\nu}\right) \, ,
\label{gravprop}
\end{equation} 
where $s$ is the Mandelstam variable, and
\beq
M_{\mu \nu}^{\, \alpha_1 \bar{\alpha}_1} \; = \; (p_{\mu}{}^{\alpha_1} p_{\nu}{}^{\bar{\alpha}_1} + p_{\nu}{}^{\alpha_1} p_{\mu}{}^{\bar{\alpha}_1} - \eta_{\mu \nu}p^{\alpha_1}\cdot p^{\bar{\alpha}_1} - \eta_{\mu \nu} V^{\alpha_1 \bar{\alpha}_1})\,,
\eeq
with a similar expression for the outgoing states $M_{\rho \sigma}^{\, \alpha_2 \bar{\alpha}_2}$. The scattering amplitude~(\ref{Agravcomplex}) becomes
\beq
{\cal A}^{z_{\alpha_{1}} \bar{z}_{\bar{\alpha}_1} z_{\alpha_2} \bar{z}_{\bar{\alpha}_2}} \; = \; \frac{2}{sM_P} \left[2(m_{\alpha_1 \bar{\alpha}_1}^4+m_{\alpha_2 \bar{\alpha}_2}^4)-2(m_{\alpha_1 \bar{\alpha}_1}^2+m_{\alpha_2 \bar{\alpha}_2}^2)(t+u)-s^2+t^2+u^2 \right] \, ,
\eeq 
where $m_{\alpha \bar{\alpha}}^2 = V_{\alpha \bar{\alpha}}$, and $s = (p_1+p_2)^2$, $t = (p_1-p_3)^2$ and $u = (p_1-p_4)^2$ are the Mandelstam invariants. 

For real fields, the energy-momentum tensor~(\ref{eq:enmomtensor1}) can be expressed as
\begin{align}
T^{\mu \nu}_{\chi, \, \xi} \; = \; \left( \partial^{\mu} \chi^{\alpha} + i \partial^{\mu} \xi^{\alpha} \right)\left( \partial^{\nu} \chi_{\alpha} - i \partial^{\nu} \xi_{\alpha} \right) -
\eta^{\mu \nu}
\left[
\frac{1}{2}\left( \partial^{\lambda} \chi^{\alpha} + i \partial^{\lambda} \xi^{\alpha} \right)\left( \partial_{\lambda} \chi_{\alpha} - i \partial_{\lambda} \xi_{\alpha} \right)-V(\chi, \xi)\right] \, ,
\end{align}
where $z^{\alpha} = \frac{1}{\sqrt{2}} (\chi^{\alpha} + i \xi^{\alpha})$ and $\bar{z}^{\bar{\alpha}} = \bar{z}^{\alpha}$.
The amplitude for the scattering of $i i \to f f$, where the initial and final states must be identical particles of the type $\{ \chi^{\alpha}, \xi^{\alpha} \}$,
then takes the form
\beq
\mathcal{A} \; = \; \frac{1}{M_P^2} M_{\mu \nu}^f \Pi^{\mu \nu \rho \sigma} M_{\rho \sigma}^i \, ,
\label{Agrav}
\eeq
where
\beq
M^i_{\mu \nu} = \frac{1}{2}(p_{1\mu} p_{2\nu} + p_{1\nu} p_{2\mu} - \eta_{\mu \nu}p_1\cdot p_2 - \eta_{\mu \nu} V_{i i}) \, ,
\eeq
and a similar expression for the outgoing states $M_{\mu \nu}^{f}$.
If we denote the mass of the initial state scalar field as $m_i$ and that of the final state as $m_f$, and we take into account factors of two for identical particles in the initial and final state, we can
evaluate Eq.~(\ref{Agrav})~\footnote{In the limit of negligible masses, this amplitude agrees with that calculated in \cite{cmov}.}
\beq
\mathcal{A} \; = \; \frac{1}{2s M_P^2}\left[2(m_i^4+m_f^4)-2(m_i^2+m_f^2)(t+u)-s^2+t^2+u^2\right]\,,
\label{realgamp}
\eeq
which in the non-relativistic limit (for the incoming states) with $t=u=m_f^2-m_i^2$ and $s=4m_i^2$, is simply
\beq
\mathcal{A} \; = \; - \frac{1}{M_P^2} \left(m_i^2+\frac{m_f^2}{2}\right) \, .
\eeq

The square of the amplitude in Eq.~(\ref{realgamp}) can be integrated to obtain a gravitational scattering cross section, $\sigma_{\rm grav}$. Using the expression
\beq
\sigma = \frac{1}{128\pi s p_i^2}\int |\mathcal{A}|^2 dt=\frac{1}{64\pi s}\frac{p_f}{p_i}\int_{-1}^{1} |\mathcal{A}|^2 d(\cos\theta)\, ,
\label{cst}
\eeq
where we have included a factor of 1/2 for identical particles in the final state, and  $p_{i,f} = \frac12 \sqrt{s-4m_{i,f}^2}$,
we obtain
\begin{align}
\sigma_{\rm grav} =  \frac{1}{960 \pi M_P^4}\frac{p_f}{p_i}&\left(  s +2(m_i^2+m_f^2)+\frac{6}{s}(m_i^4+4m_i^2 m_f^2+m_f^4)\right.\nonumber\\
&\left.-\frac{8}{s^2}m_i^2m_f^2(m_i^2+m_f^2)+\frac{56}{s^3} m_i^4 m_f^4\right)\,.
\label{csgt}
\end{align}
Note that as $\sqrt{s}=E_{CM} \rightarrow \infty$ the cross section \eqref{csgt} diverges like $s$, as expected for the (nonrenormalizable) effective theory of gravity.

\section{Testing the Conjecture with simple examples} 
\label{sec:simple}

We are now in a position to test the scalar weak gravity conjecture in some of the simple models considered in \cite{dgov} where the supergravity-induced amplitudes were calculated.
Recall that the amplitudes were found to be field-redefinition invariant only when evaluating the amplitudes in a stable vacuum.  Indeed in \cite{dgov}, it was shown that away from an extremum of the potential, the amplitudes depend on whether  we 
choose KNC or RNC.  Away from the minimum, and for minima with non-zero vacuum energy density,
the calculation of the gravitational scattering amplitudes is also incorrect as one would necessarily be forced to consider states in a non-Minkowski background. Thus here, and in the following section, we restrict our attention to calculating amplitudes in a supersymmetry preserving minimum with zero vacuum energy density.

In all of the models considered, we will assume
a maximally symmetric K\"ahler geometry with one or two chiral superfields. 
No-scale supergravity \cite{no-scale} is the natural framework for a low-energy field theory derived from string theory \cite{Witten}.
The simplest $\mathcal{N} = 1$ no-scale supergravity models were first considered in~\cite{no-scale, Ellis:1983sf}, and are characterized by the following K\"ahler potential~\cite{EKN1}:
\begin{equation} 
\label{kah1}
K \; = \; - \, 3 \alpha \ln (T + \bar{T}) \, ,
\end{equation}
where $T$ is a complex chiral field that can be identified as the volume modulus field, and $\alpha$ is a real parameter associated with the curvature of the K\"ahler manifold.  The minimal no-scale K\"ahler potential (\ref{kah1}) describes a non-compact $\frac{SU(1,1)}{U(1)}$ coset manifold. This can be generalized by including matter fields $\phi^i$ that parametrize, together with the volume modulus $T$, an $\frac{SU(N,1)}{SU(N)\times U(1)}$ coset space, 
defined by the K\"ahler potential~\cite{EKN}
\beq
\label{v0}
K \; = \; -3  \alpha \ln (T + \bar{T} - \frac{|\phi^i|^2}{3}) \, .
\eeq
We will restrict our attention to examples with $N = 1$ or 2. 
There are other generalizations based on other non-compact coset spaces, which also appear in some string models and involve multiple moduli fields.

\subsection{Single Complex Scalar Field}

\subsubsection{Massless Case}
In our first example, we consider a simple single field model with a K\"ahler potential given by
Eq.~(\ref{kah1}). 
For $W=0$, the Lagrangian is simply,
\beq
\label{eq:TLag1}
    {\cal L} = \frac{3\alpha}{(T+\bar T)^2} \partial_\mu T \partial^\mu {\bar T} 
    \, .
\end{equation}
We choose a complex background field value $w$
and define the fluctuation about that background as $T(x) = w+t(x)$. 
Using the transformation (\ref{eq:transkah1}) to KNC, which in this case can be written as 
\footnote{Note that an additional rescaling $\hat{t} \rightarrow \hat{g}_{\hat{t} \hat{\bar{t}}}^{-1/2}\hat{t}$ has been performed.}
\begin{equation}
    \label{holtrans1}
    t \; = \; \sqrt{2}v\sum_{n= 1}^{\infty} \left(\frac{\hat{t}}{\sqrt{3\alpha}} \right)^n  \, ,
\end{equation}
where $\hat{t}$ is the complex scalar field in the KNC basis and $v = \sqrt{2}\, {\rm Re}[w]$. 
In terms of real KNC fields defined by
$\hat{t} = \frac{1}{\sqrt{2}}\left(\hat{\chi} + i \hat{\xi} \right)$,
the Lagrangian up to quartic interactions is 
\begin{equation}
    \label{eq:RNLag2field}
    \mathcal{L}_{\rm KNC} = \frac{1}{2} (\partial_{\mu} \hat{\chi})^2 + \frac{1}{2} (\partial_{\mu} \hat{\xi})^2 + \frac{1}{6\alpha} \left( \hat{\chi}^2 + \hat{\xi}^2\right) \left( (\partial_{\mu} \hat{\chi})^2 + (\partial_{\mu} \hat{\xi})^2 \right) \, .
\end{equation}
The amplitudes for this simple theory have been calculated \cite{dgov} and for states with identical initial and final states they are given by
\begin{eqnarray}
    A^{\chi \chi \xi \xi}_{4} &\; = \; A^{\xi \xi \chi \chi}_{4} &\; = \; -\frac{2s}{3\alpha M_P^2} \, , \\
     A^{\chi \chi \chi \chi}_{4} &\; = \; A^{\xi \xi \xi \xi}_{4} &\; = \; 0 
     \label{4same}\, . 
\end{eqnarray}
For $W=0$, the fields are massless and $s = 4 p^2$ where $p$ is the momentum of the incoming (and outgoing) state in the center-of-mass frame. 

The gravitational scattering amplitude for this case is easily obtained from Eq.~(\ref{realgamp}) giving
\beq
{\cal A}_{\rm grav} = \frac{s}{4M_P^2}(\cos^2 \theta -1) \, ,
\label{gravamp}
\eeq
where $\theta$ is the angle between $p_1$ and $p_3$.
The amplitudes for $\chi \chi \to \chi \chi$ and $\xi \xi \to \xi \xi$ would appear to violate the sSWGC for all but forward scattering in the gravitational amplitude.
In the $\chi \chi \to \xi \xi$ and $\xi \xi \to \chi \chi$ channels, the sSWGC is satisfied provided $\alpha < 8/3$.

When we calculate the total inclusive cross section for $\chi \chi$ scattering (the cross section for $\xi \xi$ scattering is identical in this case), 
we find 
\beq
\sigma_{\rm scalar} = \sigma_{\chi \chi} = \frac{s}{72 \pi \alpha^2 M_P^4} > 
\frac{s}{480\pi M_P^4} = \sigma_{\rm gravity}\,,
\label{eq:XX}
\eeq
for $3 \alpha^2 < 20$ and all $\sigma$SWGCs are satisfied.
This theory is surely overly minimal as there
are no interactions for $T$, beyond those generated by the kinetic terms, and the potential for $T$ is completely flat, the key feature of no-scale models with the full $SU(1,1)/U(1)$ symmetry.

\subsubsection{Massive Case}
In \cite{dgov}, we also considered a model with a single chiral field, 
but with a simple quadratic superpotential
\beq
W(T) = m (T-1)^2 \,.
\eeq
The kinetic terms are still given by Eq.~(\ref{eq:RNLag2field}), but there is now a potential 
\beq
V_{\rm KNC}(\hat{\chi}, \hat{\xi})=\frac{2m^2}{27}(\hat{\chi}^2 + \hat{\xi}^2)(6 - 3\sqrt{6} \hat{\chi} + 4 \hat{\chi}^2 + 4 \hat{\xi}^2) \, ,
    \label{eq:potKNC}
\eeq
where we have expanded about the potential minimum at $v = \sqrt{2}$. At the minimum, the two scalar fields are degenerate with masses $m_{\hat{\chi}}^2 = m_{\hat{\xi}}^2 = 8m^2/9$. 
\begin{align}
    \mathcal{A}^{\chi \chi \chi \chi}_4 \; = \; \frac{32m^2}{27M_P^2} \left[-5 + 81 m^2 \left(\frac{1}{8 m^2-9 s}+\frac{1}{8 m^2-9 t}+\frac{1}{8 m^2-9 u} \right) \right]
    \, ,
    \label{cccc1}
\end{align}
\begin{align}
    \mathcal{A}^{\xi \xi \xi \xi}_4 \; = \; \frac{32m^2}{27M_P^2} \left[-5 + 9 m^2 \left(\frac{1}{8 m^2-9 s}+\frac{1}{8 m^2-9 t}+\frac{1}{8 m^2-9 u}\right) \right] \, ,
    \label{xxxx1}
\end{align} 
\begin{align}
    \mathcal{A}^{\chi \chi \xi \xi}_4 \; = \; \mathcal{A}^{\xi \xi \chi \chi}_4\; = \; \frac{2m^2}{27M_P^2} \left[-16 - \frac{9s}{m^2} + 144 m^2 \left(\frac{3}{8 m^2-9 s}+\frac{1}{8 m^2-9 t}+\frac{1}{8 m^2-9 u}\right) \right] \, ,
    \label{ccxx1}
\end{align} 
where $t = ( \frac{16}{9}m^2-\frac{s}{2})(1- \cos{\theta})$, $u = ( \frac{16}{9}m^2-\frac{s}{2})(1 + \cos{\theta})$. Analytic expressions for the cross sections can be obtained by integrating the amplitudes but are not reproduced here.

\begin{figure}[ht!]
    \centering
    \includegraphics[width=0.60\columnwidth]{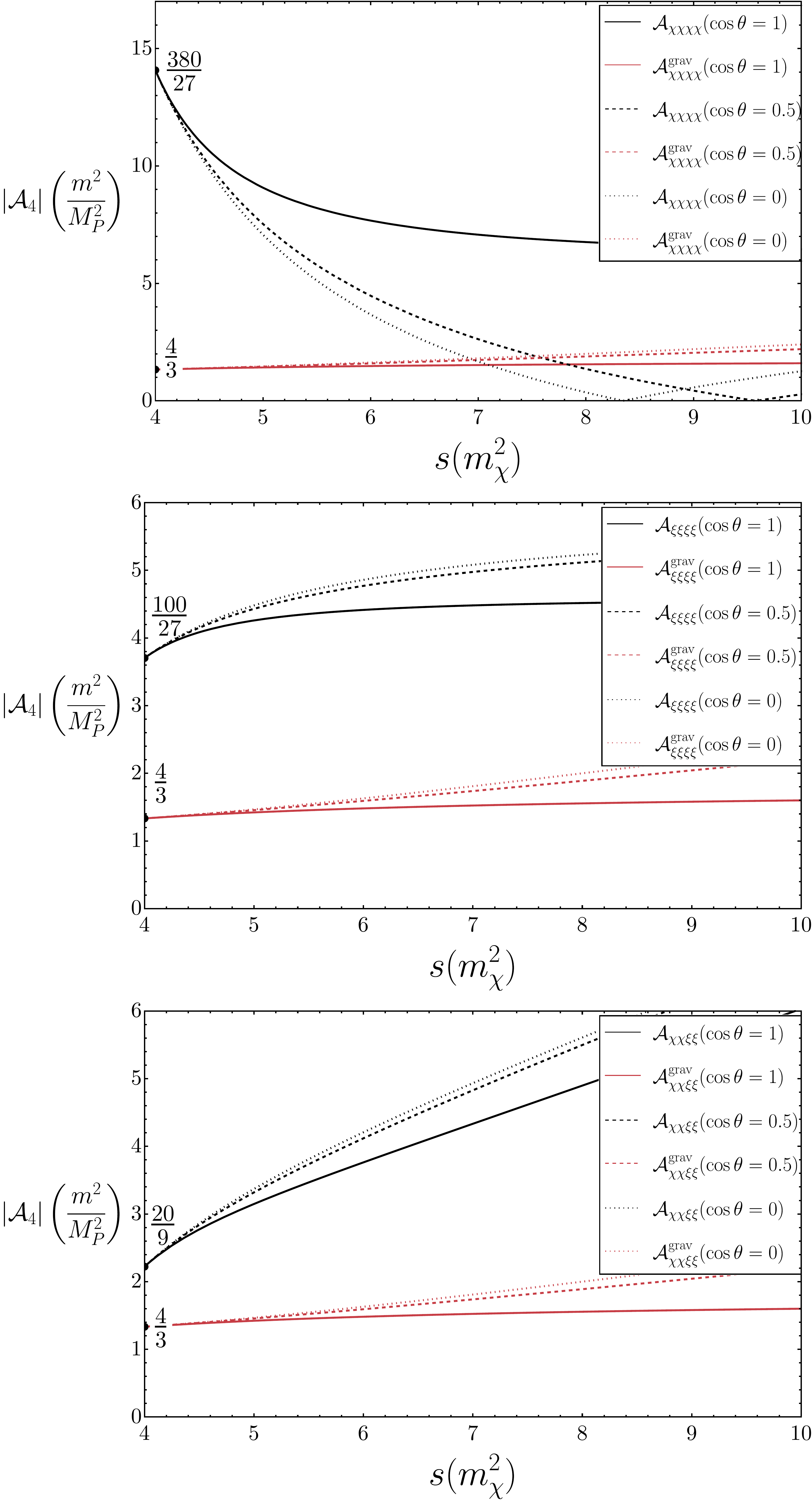}
    \caption{The scattering amplitudes given in Eqs.~\eqref{cccc1}, \eqref{xxxx1}, \eqref{ccxx1}, for three choices of the scattering angle, $\theta$. These are compared with the gravitational amplitude for the same processes. The fractions written in the limit $s\to 4 m^2$,
    correspond to the non-relativistic limit for each of the amplitudes found in \cite{dgov}. }
    \label{amp1}
\end{figure}

In the low energy limit, the gravitational scattering amplitude from Eq.~(\ref{realgamp}) is 
\beq
\mathcal{A}_{\rm grav} = - \frac{4 m^2}{3M_P^2}\,,
\eeq
which (in absolute value) is indeed smaller than the non-gravitational amplitudes which were found to be $(\frac{380}{27}, -\frac{100}{27}, -\frac{20}{9}, -\frac{20}{9})\frac{m^2}{M_P^2}$
for the $\chi \chi \to \chi \chi$, $\xi \xi \to \xi \xi$, $\chi \chi \to \xi \xi$, $\xi \xi \to \chi \chi$ channels, respectively. Since the current formulation of the sSWGC is no longer based on the formation of bound states, we may wish to extend the conjecture to higher energies and require gravitational scattering at all energies be weaker than non-gravitational scattering. 
At higher energies, we show a comparison of the scattering amplitudes in Fig.~\ref{amp1}
for each of the processes in Eqs.~\eqref{cccc1}, \eqref{xxxx1}, \eqref{ccxx1} for $\cos \theta = 0, 0.5$, and 1. These are compared with the gravitational amplitude for the same processes, as labeled. 

\begin{figure}[ht!]
    \centering   \includegraphics[width=0.9\columnwidth]{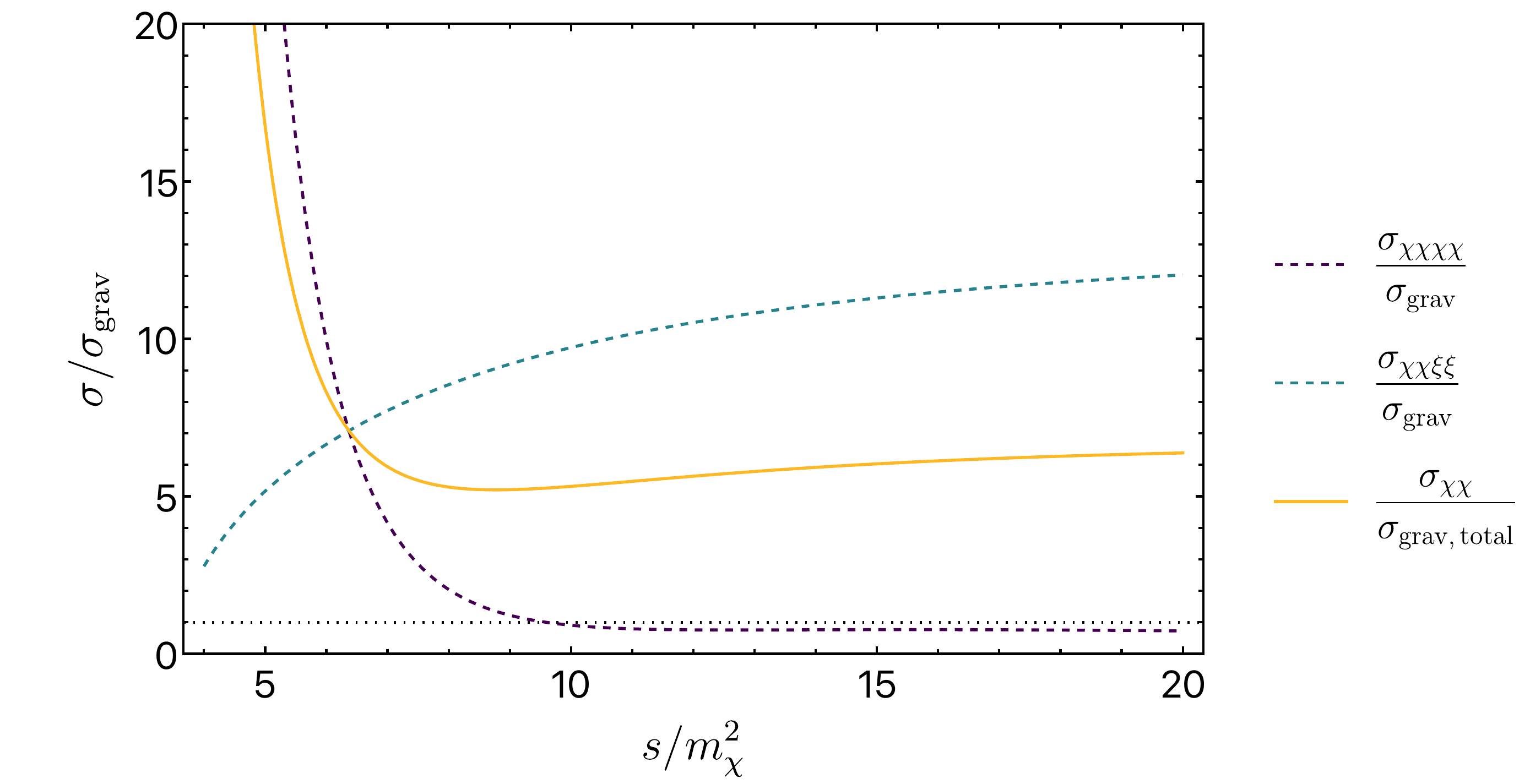}
\includegraphics[width=0.9\columnwidth]{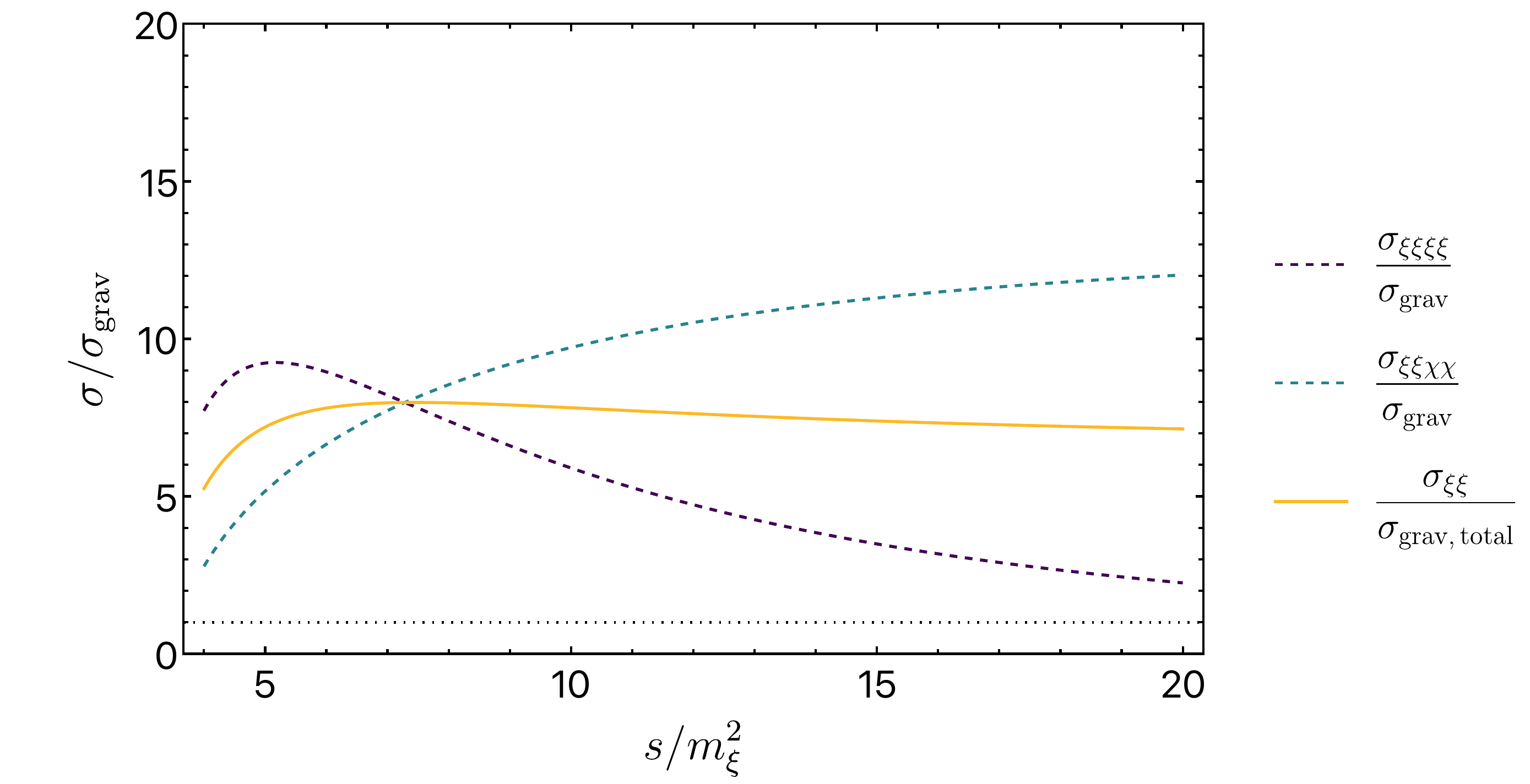}
    \caption{The ratio of non-gravitational to gravitational cross sections for (upper) $\chi \chi \to \chi \chi$ or $\xi \xi$ and (lower) $\xi \xi \to \xi \xi$ or $\chi \chi$ from the amplitudes in Eqs.~\eqref{cccc1}, \eqref{xxxx1}, \eqref{ccxx1}. Also shown is the ratio of inclusive cross sections. }
    \label{cs1}
\end{figure}

As one can see, that although the scattering amplitudes for the non-gravitational interactions are always larger than the gravitational amplitudes at low energy, at higher energy, for all but the forward scattering angle, the amplitude for $\chi \chi \to \chi \chi$ changes sign causing the amplitude to be smaller than the gravitational amplitude for some range in energy.
However, we see in Fig.~\ref{cs1}, when $|\mathcal{A}|^2$ is integrated over the scattering angle as in Eq.~(\ref{cst}), the total inclusive cross sections for both $\chi \chi$ and $\xi \xi$ scattering are greater than the gravitational cross sections. In Fig.~\ref{cs1}, 
we show separately the cross sections for $\chi \chi \to \chi \chi$ and $\chi \chi \to \xi \xi$ relative to the gravitational cross section given in Eq.~(\ref{csgt}) in the upper panel. While the former ratio is $< 1$ for $s/m_\chi^2 \gtrsim 10$, the total inclusive ratio is always greater than 1, indicating that gravity is the weakest interaction in this case and the $\sigma_{ai}$SWGC is satisfied.
The lower panel of Fig.~\ref{cs1} shows the ratio of cross sections for 
$\xi \xi \to \xi \xi$ and $\xi \xi \to \chi \chi$.

\subsection{KKLT}
\label{sec:KKLT}

It is also interesting to consider a second 
single model based on KKLT stabilization of the modulus
with superpotential {\cite{Kachru:2003aw}}
\beq
\label{sup:KKLT}
W = W_0 + B e^{-b T} \, ,
\eeq
where $W_0$ and $b > 0$ are constants. The model has a supersymmetry preserving (AdS) minimum 
which relates $W_0$ and the expectation value of $T = {\bar{T}} = v/\sqrt{2}$ at the minimum, 
\beq
W_0 \; = \; -B e^{-\frac{b v}{\sqrt{2}}} \left(1 + \frac{\sqrt{2}}{3} b v \right) \, .
\eeq
The scalar potential can be written in terms of $T$ and $\bar{T}$ as
\beq
V_{\rm KKLT} = \frac{b B \, e^{-b (T + {\bar T}) }}{3 (T + {\bar T})^2} \left( 3 (e^{b T}+e^{b {\bar T}})W_0 + B (6 + b(T +{\bar T})) \right)\,.
\label{Vkklt}
\eeq
To obtain a minimum with zero vacuum energy, we uplift the potential with \cite{Linde:2011ja}
\begin{equation}
    \Delta V \approx \frac{1}{3\sqrt{2} v^{\rm AdS}} b^2 B^2 e^{-\sqrt{2} b v^{\rm AdS}} \left(\frac{\sqrt{2} v^{\rm AdS}}{T + {\bar T}}\right)^2\,,
\end{equation}
where $v^{\rm AdS}$ is the VEV corresponding to the AdS minimum of $V_{\rm KKLT}$.
For the choice, $B=1, b=1$, and $W_0 = -10^{-12}$, the minimum occurs at $v=43.4573$ \cite{Linde:2011ja} and
the scalar masses are given by
\begin{equation}
    m_{\chi} \; = \; m_{\xi} \; \simeq \; 1.2\times 10^{-13} \, M_P \, .
\end{equation}

\begin{figure}[ht!]
    \centering
    \includegraphics[width=0.65\columnwidth]{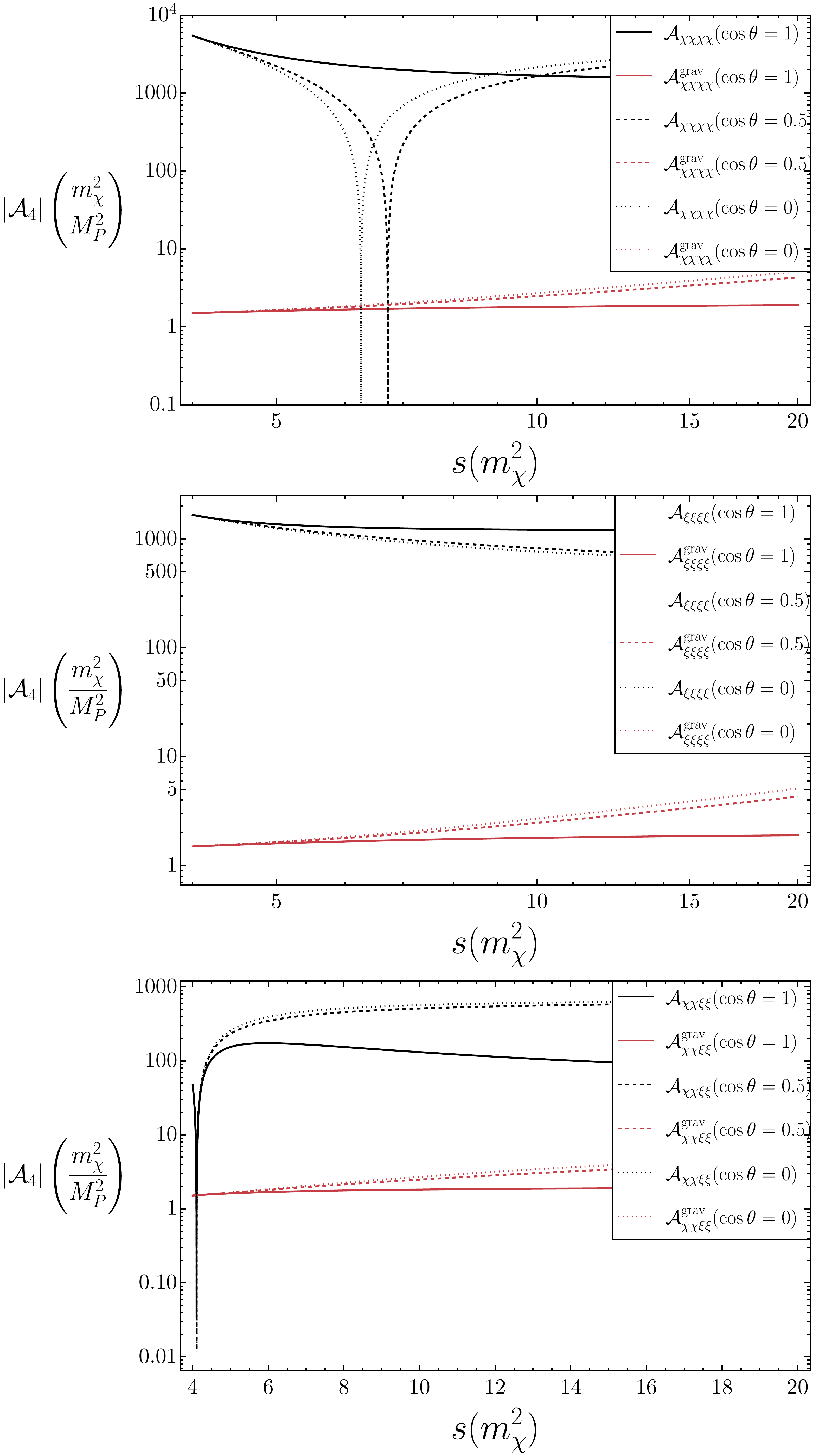}
    \caption{As in Fig.~\ref{amp1} for the uplifted KKLT model described by the scalar potential in Eq.~(\ref{Vkklt}). }
    \label{ampkklt}
\end{figure}

The expression in terms of the KNC redefined real fields, $\chi$ and $\xi$ is quite cumbersome and we rely on a numerical calculation to obtain the amplitudes.  This is shown in Fig.~\ref{ampkklt}. As in the previous example, the $\theta$-dependent amplitudes for $\chi \chi \to \chi \chi$ and in this case $\chi \chi \to \xi \xi$ change sign and therefore drop below the gravitational amplitudes for certain values of $s$.

\begin{figure}[ht!]
    \centering   \includegraphics[width=0.9\columnwidth]{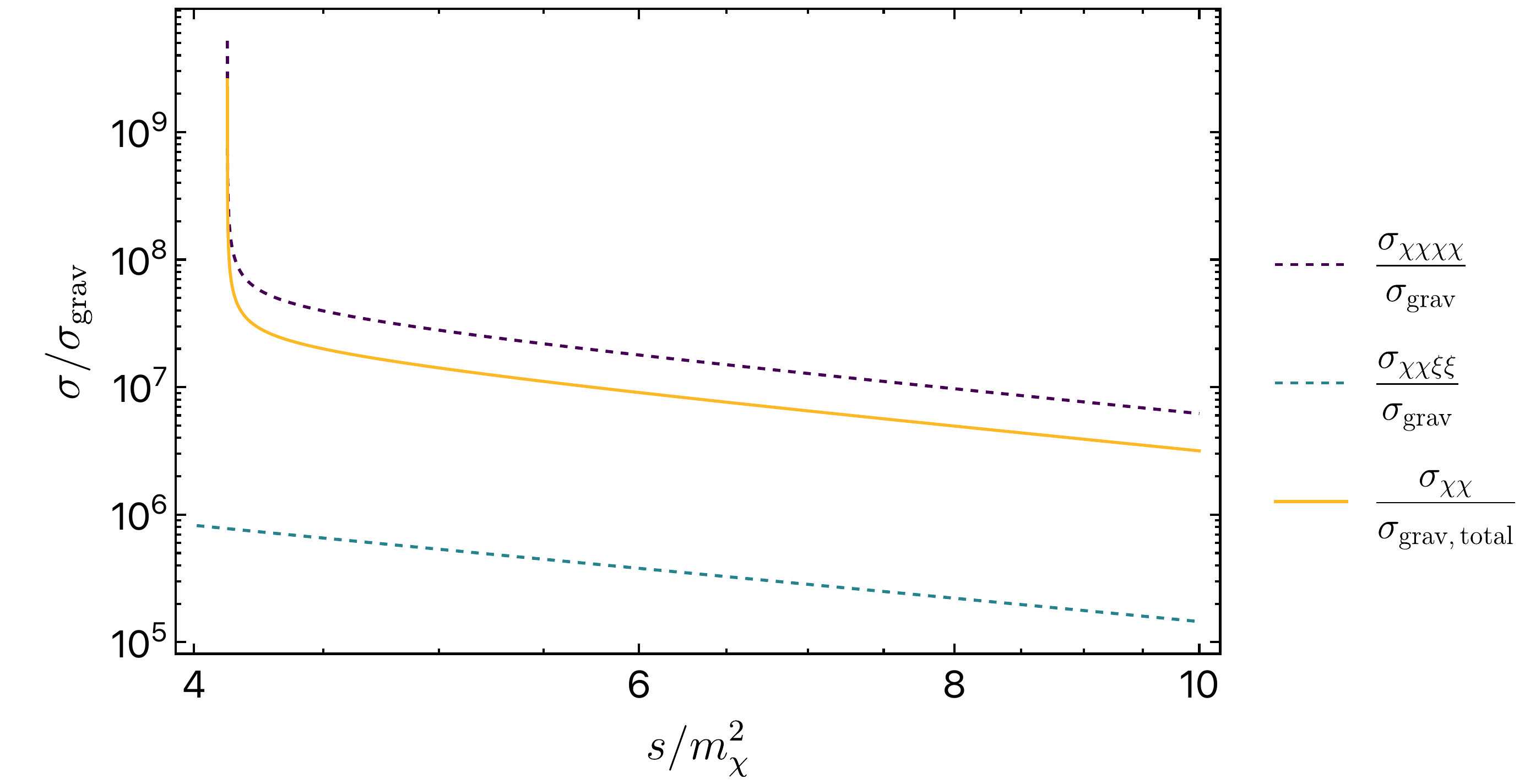}
\includegraphics[width=0.9\columnwidth]{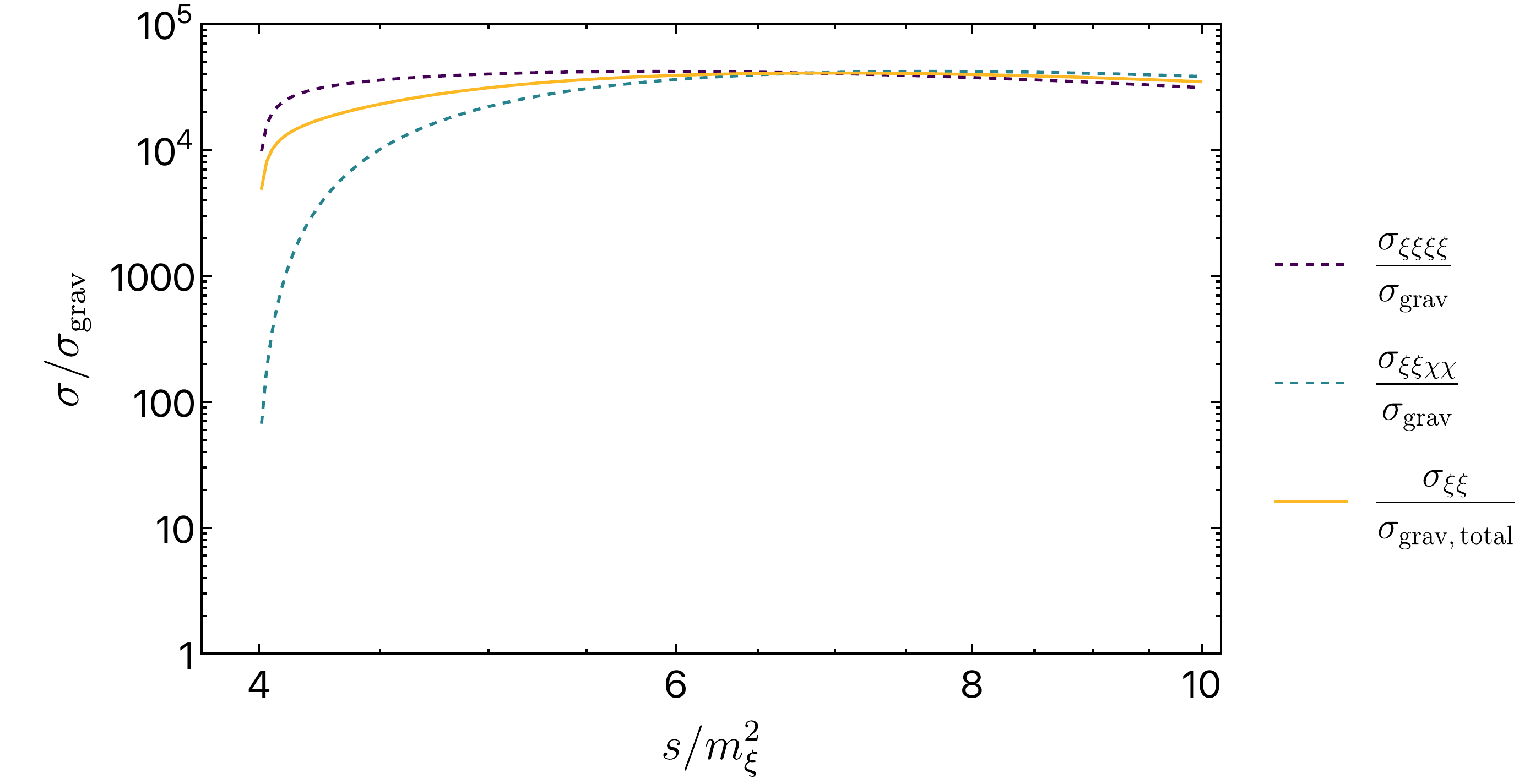}
    \caption{As in Fig.~\ref{cs1} for the uplifted 
KKLT model. }
    \label{cskklt}
\end{figure}

The corresponding scattering cross sections for the uplifted KKLT model are shown in Fig.~\ref{cskklt}. In this case, the non-gravitational cross sections are all significantly larger than the gravitational cross sections.

\subsection{Two Complex Scalar Fields}

Before moving to testing models of inflation,
we first consider a simple model involving two chiral scalar fields which may be interpreted as a modulus $T$ and matter field, $\phi$. 
In this case, we begin with the K\"ahler potential of the form given in Eq.~(\ref{v0}). The transformation to the KNC basis was given in \cite{dgov}. Once again, we expand the fields about a fixed background where $t(x)$ and $f(x)$ are the field fluctuations and $w,r$ are the field VEVs. From Eq.~(\ref{eq:transkah1}), we find for $r=0$
\begin{eqnarray}
	\label{holtrans2}
    t & =&  2 w \sum_{n=1}^{\infty} \left(\frac{\hat{t}}{\sqrt{3 \alpha}}  \right)^n \, , \\
    f & = & \sqrt{2w} \hat{f} \sum_{n = 1}^{\infty} \frac{1}{\sqrt{\alpha}^n} \left(\frac{\hat{t}}{\sqrt{3}} \right)^{n-1}\,,
    \label{holtrans3}
\end{eqnarray}
leading to kinetic terms
\begin{eqnarray}
	\label{eq:KNLag2fields2}
    \mathcal{K}_{\rm{KNC}} &= & \frac{1}{2} (\partial_{\mu} \hat{\chi})^2 + \frac{1}{2} (\partial_{\mu} \hat{\rho})^2 + \frac{1}{2} (\partial_{\mu} \hat{\xi})^2 + \frac{1}{2} (\partial_{\mu} \hat{\tau})^2 ~\nonumber \\
  &  + &\frac{1}{6\alpha} \left( ((\partial_{\mu} \hat{\chi})^2 + (\partial_{\mu} \hat{\xi})^2) (\hat{\chi}^2 + \hat{\xi}^2) + ((\partial_{\mu} \hat{\rho})^2 + (\partial_{\mu} \hat{\tau})^2) (\hat{\rho}^2 + \hat{\tau}^2)
\right.~\nonumber \\
&&\left.  
\qquad+ (\partial_{\mu} \hat{\rho} \partial^{\mu} \hat{\chi} +\partial_{\mu} \hat{\tau} \partial^{\mu} \hat{\xi})
(\hat{\rho}\hat{\chi}+\hat{\tau}\hat{\xi})+(\partial_{\mu} \hat{\rho} \partial^{\mu} \hat{\xi}-\partial_{\mu} \hat{\tau} \partial^{\mu} \hat{\chi})
(\hat{\rho}\hat{\xi}-\hat{\tau}\hat{\chi})
 \right)~\nonumber \\
  & +& \frac{1}{12\alpha} \left( ((\partial_{\mu} \hat{\chi})^2 + (\partial_{\mu} \hat{\xi})^2) (\hat{\rho}^2 + \hat{\tau}^2) + ((\partial_{\mu} \hat{\rho})^2 + (\partial_{\mu} \hat{\tau})^2) (\hat{\chi}^2 + \hat{\xi}^2)  \right)\, ,
\end{eqnarray}
where the complex fields are decomposed as $\hat{t} = \frac{1}{\sqrt{2}}(\hat{\chi} + i \hat{\xi})$ and $\hat{f} = \frac{1}{\sqrt{2}}\left(\hat{\rho} + i \hat{\tau} \right)$. This same expression will be used in computing amplitudes for the two-field models of inflation discussed in the next section.

\subsubsection{Massless case}
As in the single field example, we can 
first take $W=0$, and compare the scattering amplitudes resulting from the kinetic terms with the gravitational amplitudes. 
The relevant scattering amplitudes are \cite{dgov}
\begin{align}
    &A^{\chi \chi \xi \xi}_{4, \rm kin} \; = \; A^{\xi \xi \chi \chi}_{4, \rm kin} \; = \; A^{\rho \rho \tau\tau}_{4, \rm kin} \; = \; A^{\tau\tau \rho \rho}_{4, \rm kin} \; = \; -\frac{2s}{3\alpha M_P^2} \, , ~\nonumber\\ 
    &A^{\chi \chi \rho \rho}_{4, \rm kin} \; = \; A^{\rho \rho \chi \chi}_{4, \rm kin} \; = \; A^{\chi \chi \tau\tau}_{4, \rm kin} \; = \; A^{\tau\tau \chi \chi}_{4, \rm kin} \; = \; -\frac{s}{6\alpha M_P^2}\,.
    \label{kinamp2}
\end{align}
As in Eq.~(\ref{4same}), amplitudes with all four incoming and outgoing identical states vanish. 
The gravitational amplitudes in this case are the same as
that given in Eq.~(\ref{gravamp}). As before, the amplitudes for $\chi \chi \to \chi \chi$, $\xi \xi \to \xi \xi$, $\rho \rho \to \rho \rho$, and $\tau\tau \to \tau\tau$ violate the sSWGC for all but forward scattering in the gravitational amplitude.
In the remaining channels, the sSWGC is always satisfied provided $\alpha < 2/3$. The inclusive cross section for all four initial states ($\chi \chi$, $\xi \xi$, $\rho \rho$, and $\tau\tau$) are equal
giving 
\beq
 \sigma_{ii} = \frac{s}{64 \pi \alpha^2 M_P^4} > \frac{s}{240\pi M_P^4} = \sigma_{\rm gravity}\,,
\eeq
for $\alpha^2 < 15/4$
where $i = \chi, \xi, \rho, \sigma$ satisfying the $\sigma_{ai}$SWGC.
The additional final states increased the scalar cross section \eqref{eq:XX} by a factor of 9/8 while the gravitational cross section increased by a factor of 2. 

Before moving onto the massive case, it is appropriate to consider the role of Standard Model fields even for this toy example. 
In the case of $\chi \chi$ (or any of our initial states) scattering, the inclusive gravitational cross section (here again including only s-channel graviton exchange)
includes an additional 
contribution of $2N s/960 \pi M_P^4$, where $N = 49$ is the number of Standard Model complex scalar fields (in the Minimimal Supersymmetic Standard Model). From Eqs.~(\ref{eq:KNLag2fields2}) and (\ref{kinamp2}) we can discern that the scalar inclusive cross sections
would include a contribution of $2N s/1152 \pi M_P^4$,
giving
  \beq
 \sigma_{ii} = \frac{s}{32 \pi \alpha^2 M_P^4} \left(\frac12 + \frac{2N}{36} \right) > \frac{s}{960\pi M_P^4} (4 + 2N) = \sigma_{\rm gravity}\,,
 \label{inclsm}
\eeq    
for the $\sigma_{ai}$SWGC
where the two terms in the parenthesis on each side of the equation 
come from the toy model fields, $\phi$ and $T$ and $N$ Standard Model fields respectively. This condition is violated when $N > 33$ although the $\sigma_{a1}$SWGC is satisfied as the inequality is satisfied for a single final state.

At this point, it is worth asking what we mean by gravitational interactions. In this example, there is no superpotential and all interactions arise from the supergravity Lagrangian, in this case non-trivial kinetic interactions, and as one can see are Planck suppressed. Though Planck suppressed, we view these as a remnant of dimensional reduction
to yield the supergravity Lagrangian.  In Eq.~(\ref{inclsm}), $\sigma_{\rm gravity}$ includes only the s-channel graviton exchange. 
Additional 'scalar' interactions would arise from a non-trivial superpotential as discussed in the next subsection. 

\subsubsection{Massive case}
Since we are more interested in cases with a non-zero potential, we first consider the simple example with superpotential
\begin{equation}
    W(T,\phi) \; = \; m (T - 1)^2 + \lambda \phi^3 \, .
    \label{tphi}
\end{equation}
The resulting scalar potential in the KNC basis is
\begin{eqnarray}
\label{2fieldVKNC}
 &&V_{\rm{KNC}} = \frac{4}{9} m^2 (\tilde{\chi }^2+\tilde{\xi }^2)+\frac{9}{4} \lambda ^2(\tilde{\rho }^2 +\tilde{\tau}^2)^2-\frac{2}{3} \sqrt{\frac{2}{3}} m^2 (\tilde{\xi }^2 \tilde{\chi }+\tilde{\chi }^3)~\nonumber\\
 &&\qquad\qquad+\frac{8}{27} m^2 (\tilde{\chi }^2+\tilde{\xi }^2)^2+\frac{4}{27} m^2 (\tilde{\chi }^2+\tilde{\xi }^2)(\tilde{\rho }^2+\tilde{\tau}^2)\,.
\end{eqnarray}
The fields $\chi$ and $\xi$ are again massive with 
 $m_{\hat{\chi}}^2 = m_{\hat{\xi}}^2 = 8m^2/9$ as in the previous example. However, in the absence of a bilinear term for $\phi$, or supersymmetry breaking, the matter fields, $\rho$ and $\tau$ remain massless. 
 
 For this example, the amplitudes for 
 $\chi \chi \to \chi \chi$, $\xi \xi \to \xi \xi$ and $\chi \chi \to \xi \xi$ are identical to those given in Eqs.~\eqref{cccc1},\eqref{xxxx1},\eqref{ccxx1}.
 The additional relevant amplitudes including kinetic contributions for $\alpha = 1$ are
\begin{align}
    \mathcal{A}^{\chi \chi \rho \rho}_4 \; = \; \mathcal{A}^{ \rho \rho \chi \chi}_4 \; = \; \mathcal{A}^{\chi \chi \tau\tau}_4 \; = \; \mathcal{A}^{ \tau\tau \chi \chi}_4 \; =& \; -\frac{8m^2}{27M_P^2} - \frac{s}{6M_P^2} \,,
    \label{ac1}
\end{align} 
\begin{align}
    \mathcal{A}^{\xi \xi \rho \rho}_4 \; = \; \mathcal{A}^{ \rho \rho \xi \xi}_4 \; = \; \mathcal{A}^{\xi \xi \tau\tau}_4 \; = \; \mathcal{A}^{ \tau\tau \xi \xi}_4 \; = &\; -\frac{8m^2}{27M_P^2} - \frac{s}{6M_P^2} \,,
        \label{ax1}
\end{align} 
\begin{align}
    \mathcal{A}^{\rho \rho \rho \rho}_4 \; = \;  \mathcal{A}^{\tau\tau \tau\tau}_4 \; = &\; -54 \lambda^2 \,,
        \label{ar1}
\end{align} 
\begin{align}
    \mathcal{A}^{\rho \rho \tau\tau}_4 \; = \;  \mathcal{A}^{\tau\tau \rho \rho}_4 \; = &\; -\frac{2s}{3M_P^2} - 18 \lambda^2 \, .
        \label{ar2}
\end{align} 
These can be compared to the gravitational amplitude in Eq.~(\ref{gravamp}). 

\begin{figure}[ht!]
    \centering   \includegraphics[width=0.9\columnwidth]{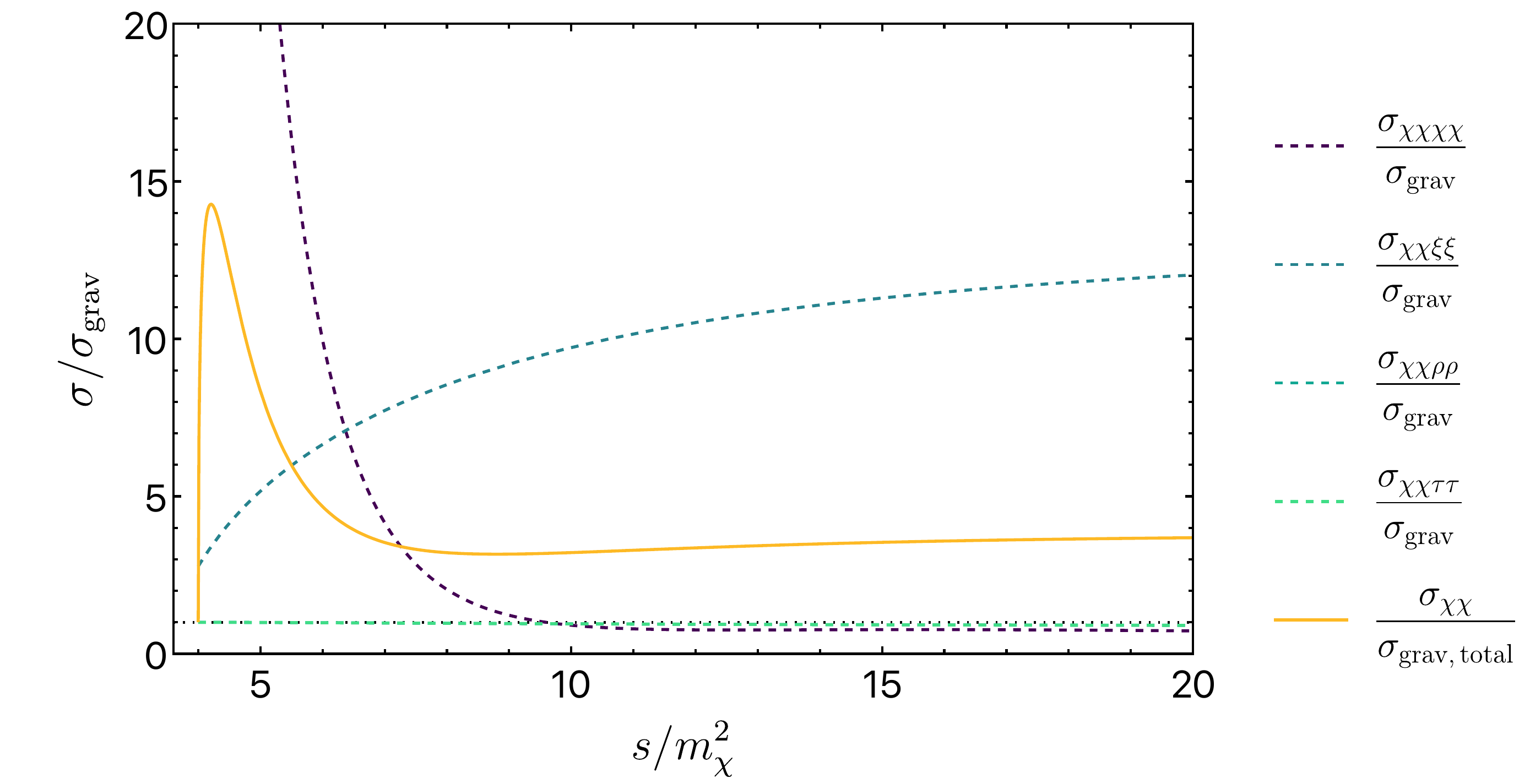}
\includegraphics[width=0.9\columnwidth]{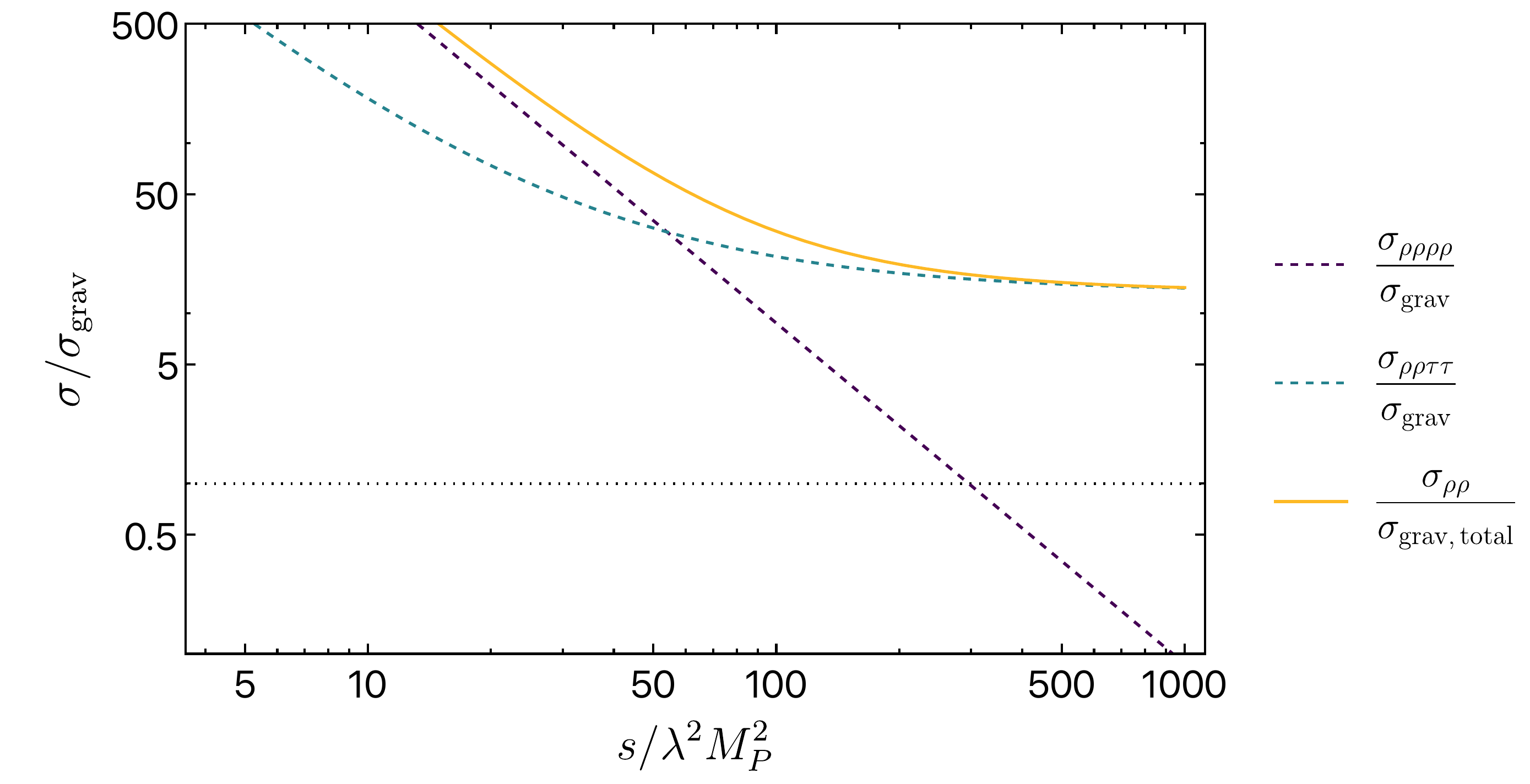}
    \caption{As in Fig.~\ref{cs1} for the two-chiral field model with superpotential (\ref{tphi}).}
    \label{cs2}
\end{figure}

In Fig.~\ref{cs2}, we show the scattering cross sections for $\chi \chi$ (upper panel) and $\rho \rho$ (lower panel) initial states. As one can see, for $\chi \chi$ initial states, the cross sections to $\chi \chi$ and $\xi \xi$ are the same as in the previous section, and the cross section to $\rho \rho$ and $\tau\tau$ are small in comparison. Since there are twice as many final states, at large $s$, the gravitational cross section is twice as large compared to the two-field example and the ratio of scalar to gravity cross sections is down by a factor of 2 as seen by comparing Figs.~\ref{cs1} and \ref{cs2}.

The cross sections for $\rho \rho$ (or $\tau \tau$) to any final state are easily obtained from Eqs.~(\ref{ac1}) - (\ref{ar2}). They are
\beq
\sigma_{\chi \chi,\xi\xi} = \frac{s}{1152\pi M_P^4}\left(1+\frac{2m_\chi^2}{s}\right)^2\sqrt{1-\frac{4m_\chi^2}{s}} \, ,
\eeq
for $\rho \rho \to \chi \chi$, $\rho \rho \to \xi \xi$.
Also,
\beq
\sigma_{\rho\rho} = \frac{729 \lambda^4}{8 \pi s} \, ,
\eeq
for $\rho\rho \to \rho \rho$, and
\beq
\sigma_{\tau\tau} = \frac{s}{72\pi M_P^4} \left(1+27\lambda^2\frac{M_P^2}{s} \right)^2 \, ,
\eeq
for $\rho\rho \to \tau\tau$. These should be compared with the gravitational cross sections which at large $s$ are simply
\beq
\sigma_{\rm grav} = \frac{s}{960\pi M_P^4} \, ,
\eeq
for each channel. 
The cross section for $\tau\tau$ initial states is the same.  At large $s$, the ratio of the scalar cross sections to gravity for $\rho \rho \to \chi \chi$ is simply 5/6.
While the ratio of the $\rho \rho \to \rho \rho$ cross sections is large at small $s$, the ratio is $<1$ for $s \gtrsim 300 \lambda^2 M_P^2$. However at large $s$, the ratio of the $\rho \rho \to \tau\tau$ cross sections is 40/3 and hence the scalar cross section always exceeds the gravitational cross section as seen in Fig.~\ref{cs2}.

Before concluding this section, we come back to the question of which interactions are gravitational. Only the processes $\rho \rho \to \rho \rho$ and $\rho \rho \to \tau \tau$ (and similarly for $\tau \tau$ initial states), have interactions which
are not Planck suppressed. As these cross sections fall with $1/s$ as $s$ increases, while the gravitational cross sections increase with $s$.
Should the conjecture be applied only at low $s$?

\section{Confronting Inflation in No-scale Supergravity}
\label{sec:infl}

Perhaps a more interesting confrontation of  supergravity models with the sSWGC arise in two-field no-scale models \cite{eno6,eno7,enov1,FeKR,others,building} which lead to Starobinsky inflation \cite{Staro}.  In \cite{Gonzalo:2019gjp,Benakli:2020pkm}, it was argued that single field Starobinsky inflationary models must be part of the swampland as the gravitational scattering amplitudes exceeded the inflaton self-scattering amplitudes at large background field values.  At the minimum, however, gravity was found to be the weakest force.
As we have argued above and in \cite{dgov}, away from the minimum, it is not possible to calculate accurate scattering amplitudes while neglecting the effect of the tadpole contributions. Furthermore, at large field values, one can not simply calculate the gravitational amplitude using the propagator in Eq.~(\ref{gravprop})
for which the gravitational fluctuations, $h_{\mu\nu}$ are expanded in a Minkowski background.

Nevertheless, it remains an interesting question whether or not the interactions of the inflaton and related fields derived from no-scale supergravity satisfy the conjecture. In no-scale supergravity, at least two chiral fields are necessary to obtain a direction in field space which yields Starobinsky inflation \cite{eno7}. 
Starting with a K\"ahler potential of the form (\ref{v0}),
there are many choices for the superpotential, related by the SU(2,1)/SU(2)$\times$U(1) symmetry, which result in the same scalar potential in a given direction \cite{eno7,enov1}. Even the identity of the inflaton (as either $T$ or $\phi$) is arbitrary up to a field redefinition.  This is perhaps not surprising as (\ref{v0}) can also be written 
as 
\beq
K = - 3 \alpha \ln \left( 1 - \frac{y_1^2}{3} - \frac{y_2^2}{3} \right)\,,
\eeq
up to a K\"ahler transformation of the superpotential. For the inflationary models considered here, we restrict our analysis to $\alpha = 1$.

We will consider two common examples. One in which Re\,$\phi$ is identified with the inflaton and
the superpotential has a Wess-Zumino-like form \cite{eno6}
\beq
W =  M \left( \frac12 \phi^2 - \frac{1}{3\sqrt{3}} \phi^3 \right) \,,
\label{wzinfl}
\eeq
where $M$ is the inflaton mass at the minimum.
With another choice of superpotential \cite{Cecotti},
\beq
W = \sqrt{3} M \phi \left( T - \frac12 \right)\, ,
\label{cinfl}
\eeq
 Re\,$T$ is identified with the inflaton and also leads to Starobinsky-like inflation. 
 The mass scale, $M$, is fixed by the amplitude of CMB anisotropies to be $M \approx 1.5 \times 10^{-5} M_P$.
We discuss both of these possibilities in the next subsections.  As we will see, successful inflation in either case will require a modification of the K\"ahler potential so that one of the two fields is stabilized during inflation. This is most easily done by adding a quartic term in the argument of the logarithm in the K\"ahler potential \cite{EKN3,eno7,EGNO4} or by imposing a superfield constraint eliminating the corresponding scalar field.

In the absence of a stabilization term, the transformation to the KNC basis and the kinetic part of the Lagrangian is given by Eqs.~(\ref{holtrans2}), (\ref{holtrans3}) and (\ref{eq:KNLag2fields2}), respectively. Similarly, the kinetic contributions to the scattering amplitudes are given by Eq.~(\ref{kinamp2}). We discuss the modifications to the transformations and kinetic terms due to stabilization for each of the two models in their respective subsections. 
Because the mass spectra of the two models differ, the derived scattering amplitudes and cross sections differ and we discuss each of the two models separately. 

\subsection{Wess-Zumino-Like Inflation}

Combining the K\"ahler potential (\ref{v0}) with the superpotential (\ref{wzinfl}), the scalar potential along the direction
$T = {\bar T} = \frac12$, Im\,$\phi = 0$,
is of the Starobinsky form 
\beq
V = \frac34 M^2 \left( 1- e^{-\sqrt{2/3}x} \right)^2 \, ,
\eeq
where along this direction, we define
the canonical field $\tilde \phi  = \sqrt{3} \tanh^{-1} (\phi/\sqrt{3})$, and $\tilde \phi = (x+iy)/\sqrt{2}$. Indeed, inflation only occurs if the volume modulus is stabilized (Im\,$\phi = 0$ minimizes the potential). 
One possibility for the stabilization of the modulus is achieved by adding quartic couplings to the argument of the K\"ahler potential \cite{EKN3,eno7,EGNO4} of the form 
\begin{equation}
\label{kstab2}
K \; = \; -3 \, \ln \left(T+{\bar T}-\frac{|\phi|^{2}}{3}+\frac{\left(T+{\bar T}-1\right)^{4}+d\left(T-{\bar T}\right)^{4}}{\Lambda_T^{2}}\right) \, ,
\end{equation}
where $d$ is a constant and $\Lambda_T$ is a mass scale smaller than the Planck scale $M_P$. The first quartic term
stabilizes the real part and leads to ${\rm{Re}} \, T = \frac12$, and 
the second term stabilizes the potential in the imaginary direction of $T$ with ${\rm{Im}} \, T = 0$. For $d = 1$, the strength of the stabilization in the real and imaginary directions is equal. When supersymmetry is broken, stabilization provides masses for $T$
\begin{equation}
m_{{\rm{Re}} \, T}^2 = 144 \, \frac{m_{3/2}^2}{\Lambda_T^2}; \qquad m_{{\rm{Im}} \, T}^2 = 144 \, d \, \frac{m_{3/2}^2}{\Lambda_T^2},
\label{mT}
\end{equation}
which are hierarchically larger than the gravitino mass if $\Lambda_T\ll 1 $.
During inflation supersymmetry is broken 
as $W\ne 0$. At the minimum, supersymmetry is most simply broken by adding a constant,
$m_{3/2}$, to the superpotential. We assume $m_{3/2} = 1$ TeV. 

The stabilization terms require a modification of the transformation of $t$ in Eq.~(\ref{holtrans2}) to express the Lagrangian in terms of KNC. To fourth order in ${\hat t}$ this is 
\beq
\Delta t = - \frac{16w^4}{\Lambda_T} \left[ 4(1-d)\left(\frac{{\hat t}}{\sqrt{3}}\right)^3 + (5-3d)\left(\frac{{\hat t}}{\sqrt{3}}\right)^4 \right] \, .
\eeq
Note that for $d=1$, the lowest order correction is quartic. 
The effect of the modified transformations induces additional kinetic terms to be 
added to Eq.~(\ref{eq:KNLag2fields2})
\beq
\label{eq:KNLag2fields3wz}
 \Delta   \mathcal{K}_{\rm{KNC}} =  \frac{2(1+d)}{\Lambda_T^2}\left(\chi^2 + \xi^2 \right) \left( (\partial_\mu \chi)^2 +(\partial_\mu \xi)^2 \right)\,,
\eeq

\begin{figure}[ht!]
    \centering
\includegraphics[width=1\columnwidth]{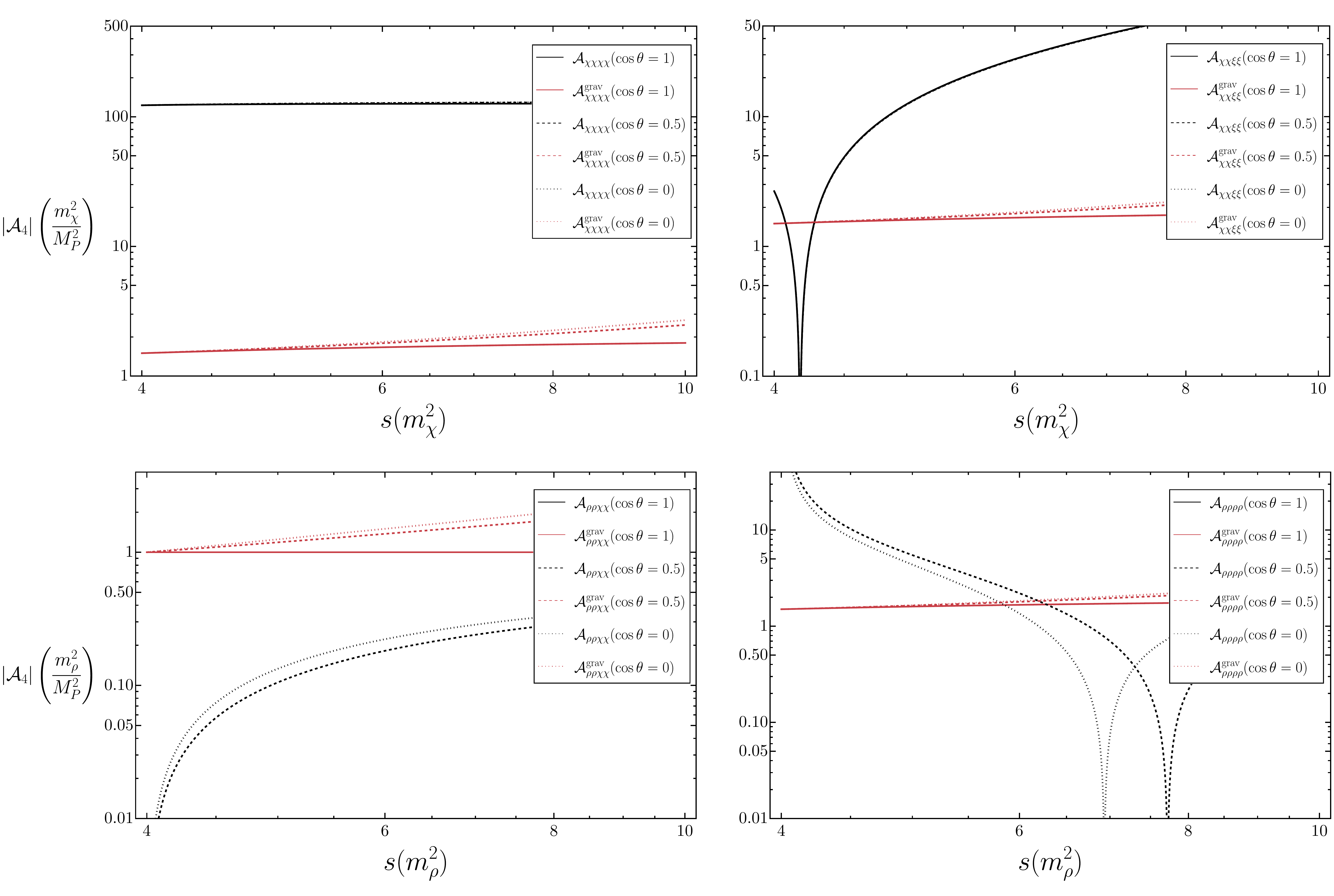}
    \caption{The scattering amplitudes for the Wess-Zumino-like inflationary model for three choices of the scattering angle with $d=1$ and $\lambda_T=M_P$. These are compared with the gravitational amplitude for the same processes. Shown are the amplitudes for $\chi \chi \to \chi \chi$ (upper left), $\chi \chi \to \xi \xi$ (upper right), $\rho \rho \to \chi \chi$ (lower left), and $\rho \rho \to \rho \rho$ (lower right).  }
    \label{ampi1}
\end{figure}

\begin{figure}[ht!]
    \centering   \includegraphics[width=0.9\columnwidth]{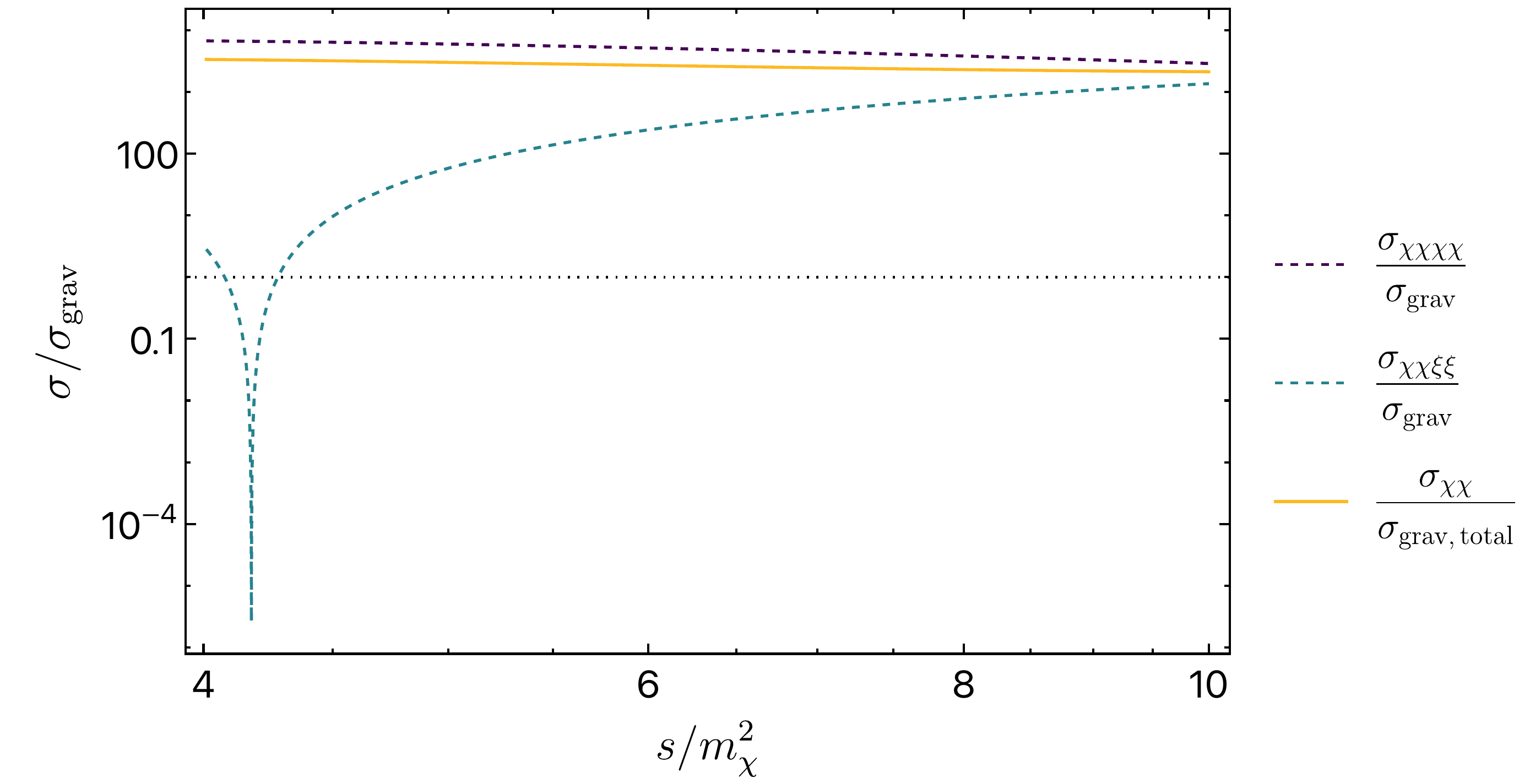}
\includegraphics[width=0.9\columnwidth]{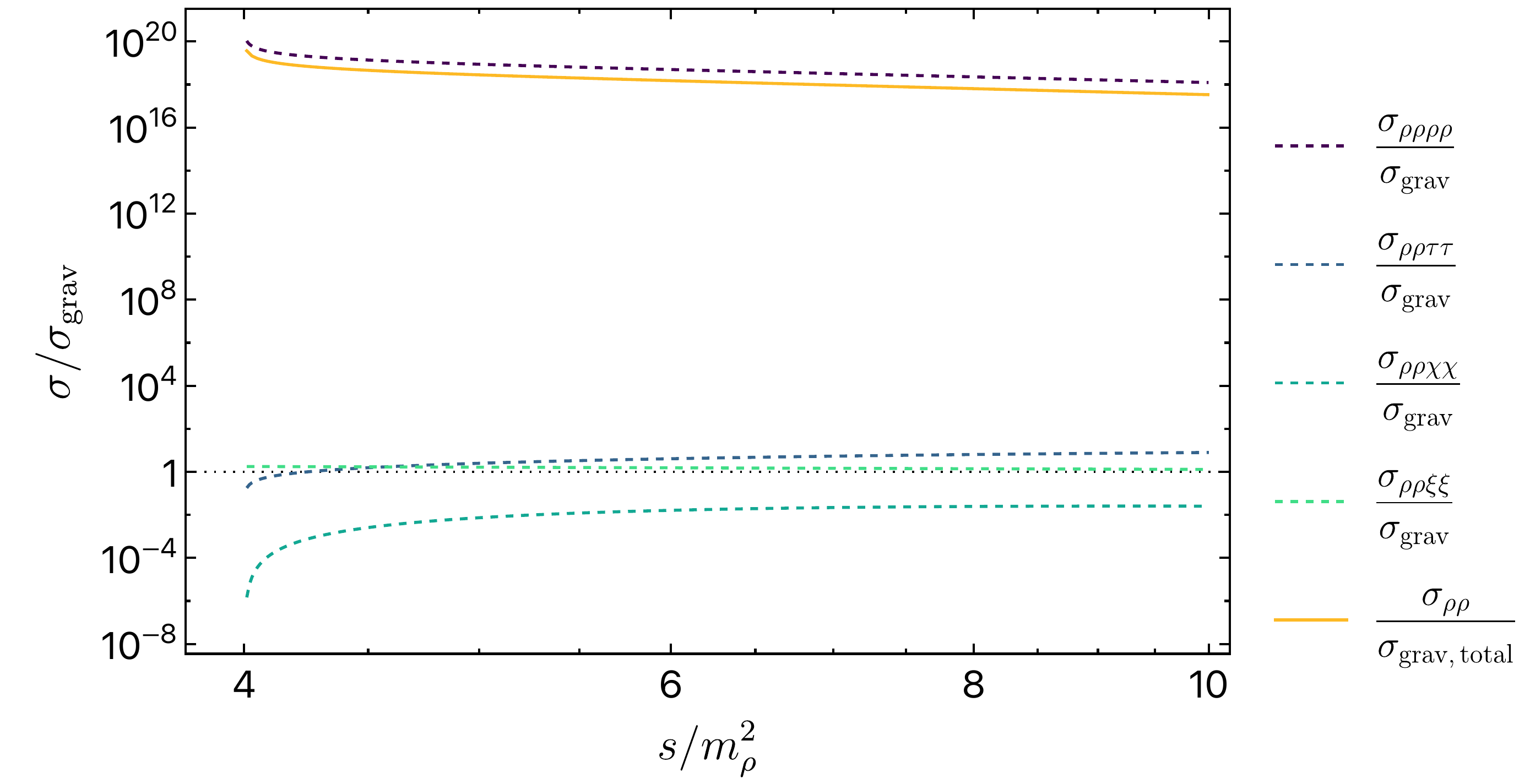}
    \caption{The ratio of the scalar scattering cross section relative to the gravitational cross section as a function of $s/m^2$ for the Wess-Zumino-like inflationary model. In the upper panel, the ratio of cross sections for $\chi \chi \to \chi \chi$ and $\xi \xi$ as well as the inclusive cross section ratio are shown. In the lower panel, the ratio of cross sections for $\rho \rho \to \rho \rho$, $\tau \tau, \chi \chi$, and $\xi \xi$, as well as the inclusive cross section are shown. }
    \label{csinf1}
\end{figure}

At the potential minimum, as in the previous (two-field) example, our canonical fields are $\hat{t} = \frac{1}{\sqrt{2}}(\hat{\chi} + i \hat{\xi})$ and $\hat{f} = \frac{1}{\sqrt{2}}\left(\hat{\rho} + i \hat{\tau} \right)$. The masses of the fluctuations about the minimum at $T = 1/2$ and $\phi = 0$ are $m_\chi = 4 \sqrt{3} m_{3/2}/\Lambda_T$, $m_\xi = 4 \sqrt{3} d m_{3/2}/\Lambda_T$, and $m_\rho = m_\tau = M$.
The amplitudes for $\chi \chi \to \chi \chi$, $\chi \chi \to \xi \xi$ and $\rho \rho \to \rho \rho$,  $\rho \rho \to \chi \chi$ scattering are shown in Fig.~\ref{ampi1} for $d=1$ and $\Lambda_T = M_P$. The partial and inclusive cross sections for these initial states are shown in Fig.~\ref{csinf1}. 
The processes $\chi \chi \to \rho \rho$ and $\chi \chi\to \tau \tau$ are kinematically forbidden unless $s$ is sufficiently large (far beyond what is shown in these plots). 
The amplitudes and cross sections for 
$\rho \rho \to \xi \xi$ and $\rho \rho \to \tau \tau$ as well as $\xi \xi$ and $\tau \tau$ scattering are similar and therefore not shown.

For the $\chi \chi$ initial state (effectively the volume modulus $T$),
once again, we see that the amplitude in one channel ($\chi \chi \to \xi \xi$) becomes smaller than the gravitational amplitude since it changes sign. The integrated cross section for this process also falls below the gravitational cross section and has a minimum at $s \simeq 4.18 m_\chi^2$. However, the $\chi \chi \to \chi \chi$ cross section is large for all $s$, dominated by the quartic contact term,  and the total inclusive cross section is  larger than the gravitational cross section as seen in the top panel of Fig.~\ref{csinf1}.
Inflaton scattering is shown here as $\rho \rho$ scattering. While several of the final state channels have cross sections below the gravitational cross section, the total scattering cross section is dominated by $\rho \rho \to \rho \rho$ scattering. This channel receives large contributions from $t$ and $u$ channel $\chi$ exchange.
Since $m_\chi \ll m_\rho$, this cross section is many orders of magnitude larger than the gravitational cross section as shown in bottom panel of  Fig.~\ref{csinf1}.

As discussed earlier, near the minimum, we can be confident that the scattering amplitudes are given by those at the minimum. We have also argued that when far from a minimum, the amplitudes should be calculated at the appropriate extremum determined by the Newton tangent method \cite{dnps}. Thus for most if not all of the inflationary trajectory, the appropriate extremum to calculate a scattering
amplitude is at $T = 1/2$ and $\phi \to \sqrt{3}$ corresponding to $x \to \infty$. In terms of the KNC fields, this corresponds to $w = \frac{1}{\sqrt{2}}(v + i u)$ with $v=  \frac{1}{\sqrt{2}}$ and $u=0$ and $r = \frac{1}{\sqrt{2}}(p + i t)$ with $p = \sqrt{6}$ and $t = 0$. This is an extreme limit of the field space manifold. Indeed, the argument of the logarithm of the K\"ahler potential vanishes in this limit. 
We note that in the absence of the stabilization term this limit would not correspond to an extremum as there would be no minimum in the $T$ direction. 
For any finite value of $\Lambda_T$, the minimum occurs at approximately $v = \frac{1}{\sqrt{2}}$ - the approximation improves as $\Lambda_T \to 0$. 
Thus, any attempt at a calculation in this limit is 
approximate since $v =\frac{1}{\sqrt{2}}$ is not the exact minimum, and therefore there are residual tadpoles which should (but have not) been accounted for. Numerically, instabilities may occur as the limit in several divergent quantities may not be well defined.
Finally, in this limit, the background space-time metric
is no longer Minkowski but rather de Sitter. In principle the scattering amplitudes should be recomputed in a de Sitter background and in particular the gravitational amplitude would be modified when using the correct graviton propagator. We have not considered these effects. 

\begin{figure}[ht!]
    \centering   \includegraphics[width=0.9\columnwidth]{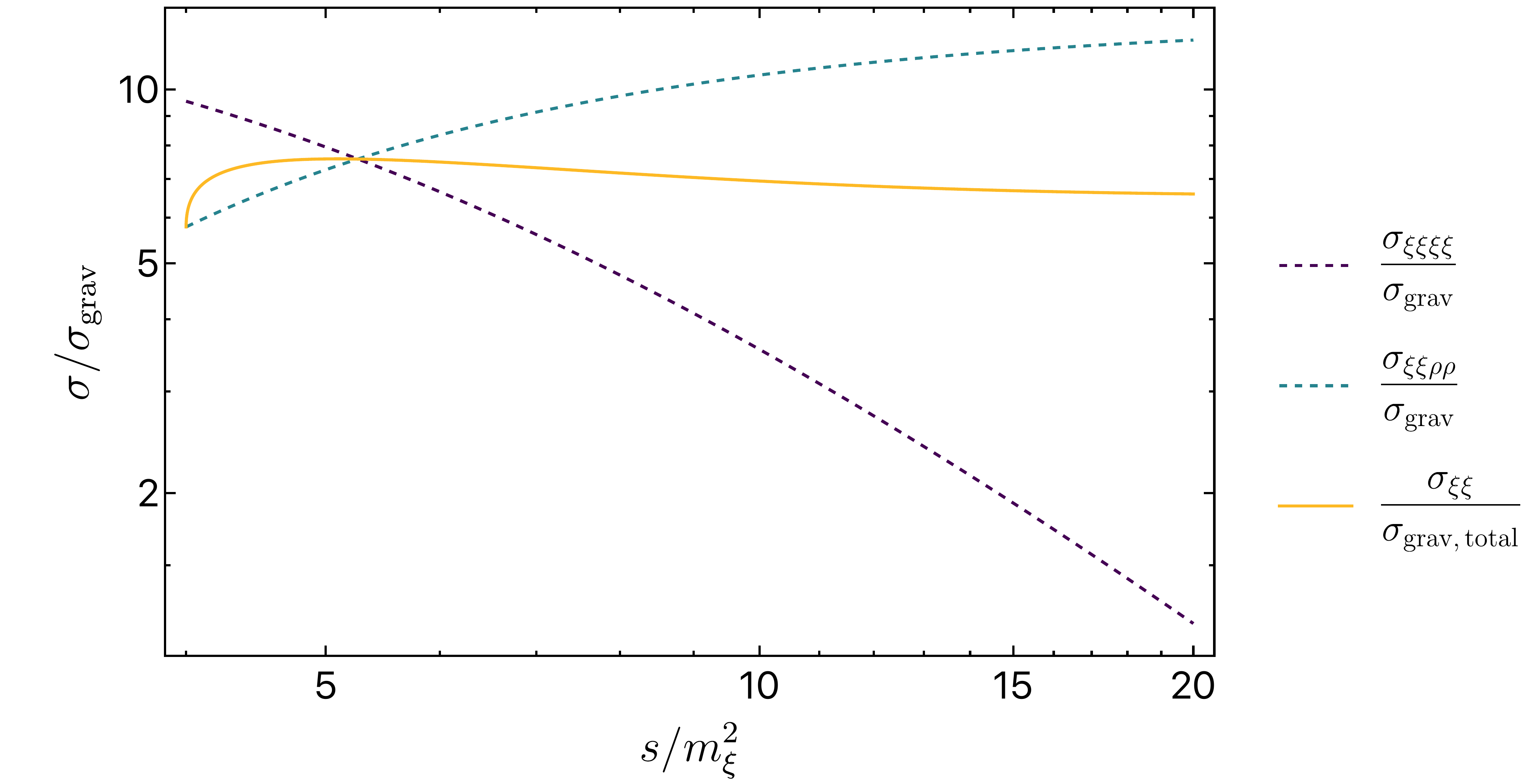}
\includegraphics[width=0.9\columnwidth]{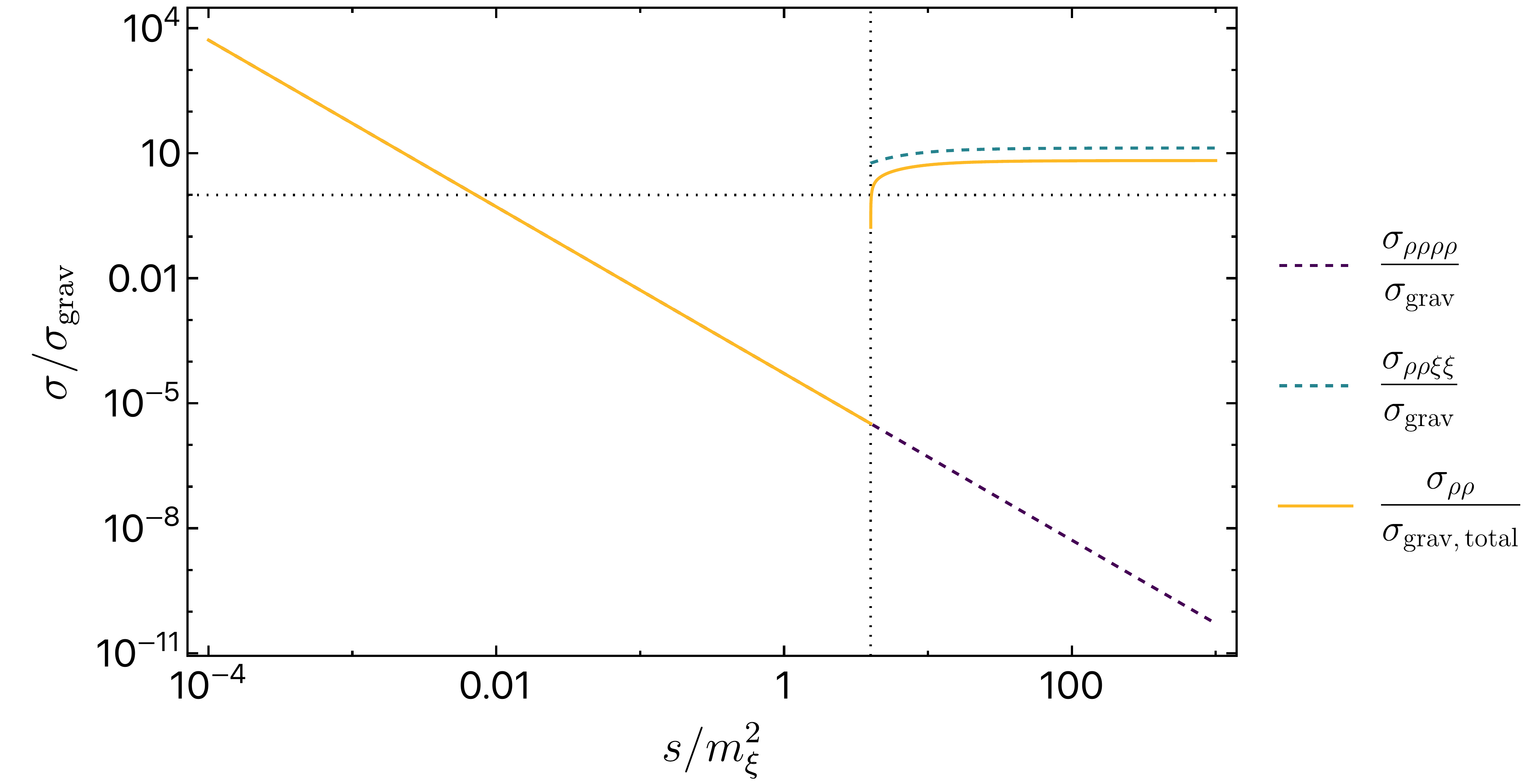}
    \caption{As in Fig.~\ref{csinf1} for scattering when the background field values take the limit, $v \rightarrow \frac{1}{\sqrt{2}}$ and $p \rightarrow \sqrt{6}$.}
    \label{wzstab}
\end{figure}

In the limit $v \rightarrow \frac{1}{\sqrt{2}}$ and $p \rightarrow \sqrt{6}$, 
the masses of $\chi$ and $\tau$ diverge, leaving us with two `light' scalars: $\xi$ with mass $m_\xi = M \sqrt{12 M_P^2 + \Lambda_T^2}/\Lambda_T$ and the inflaton $\rho$, which is massless in this limit.  The scattering cross sections compared to gravity are shown in Fig.~\ref{wzstab} with $d = 1$ and $\Lambda_T = M_P$. Final states for $\xi \xi$ scattering include both $\xi \xi$ and $\rho \rho$ for any center of mass energy and the cross section is always larger than the gravitational cross section.
In the case of $\rho \rho$ scattering, at low energy,
only $\rho \rho$ final states are allowed. As this cross section decreases with $s$, for $s \gtrsim 0.01 m_\xi^2$, the scalar cross section violates the $\sigma_{ai}$SWGC until $s>4 m_\xi^2$ when $\rho \rho  \to \xi \xi$ scattering becomes kinematically viable. In this case, even $\sigma_{a1}$SWGC is violated in the energy interval $0.01 < s/4m_\xi^2 < 4$. However, because $\xi \xi \to \rho \rho$ scattering is greater than its gravitational counterpart, the $\sigma_{1i}$SWGC remains satisfied. 

Note that in this model during the inflationary epoch, along the potential plateau, the kinetic terms approach a diverging pole, which are also known as pole-inflation \cite{Galante:2014ifa}, \cite{Broy:2015qna}. It was argued in \cite{Scalisi:2018eaz}
that such models generically violate the Swampland Distance Conjecture (SDC) \cite{Ooguri:2006in}. It is tempting to speculate that the violation of $\sigma_{ai}$SWGC and $\sigma_{a1}$SWGC in some energy range could be related to the 
violation of the SDC. 

\subsection{Cecotti-Like Inflation}

Alternatively, the volume modulus, $T$, may play the role of the inflaton if we combine the K\"ahler potential (\ref{v0})
with the superpotential (\ref{cinfl}). In this case, the Starobinsky potential is found along the direction $\phi = 0$ and Im$T=0$ for a canonical field, $x$ defined by $T = \frac12 e^{\sqrt{2/3}x}$.

Once again, inflationary dynamics require stabilization, in this case for $\phi$. 
Quartic stabilization is again sufficient and can be achieved with the following K\"ahler potential
\begin{equation}
    \label{kahpot2}
    K \; = \; -3 \alpha \ln (T + \bar{T} - \frac{|\phi|^2}{3} + \frac{|\phi|^4}{\Lambda_{\phi}^2}) \, .
\end{equation}
In this case, there are no changes to the field redefinitions given in Eq.~(\ref{holtrans2}) and (\ref{holtrans3}), and the kinetic terms receive a correction given by
\beq
\label{eq:KNLag2fields3c}
  \Delta  \mathcal{K}_{\rm{KNC}} =  \frac{3}{\Lambda_\phi^2}\left(\rho^2 + \tau^2 \right) \left( (\partial_\mu \rho)^2 +(\partial_\mu \tau)^2 \right)\,.
\eeq

\begin{figure}[ht!]
    \centering
    \includegraphics[width=1\columnwidth]{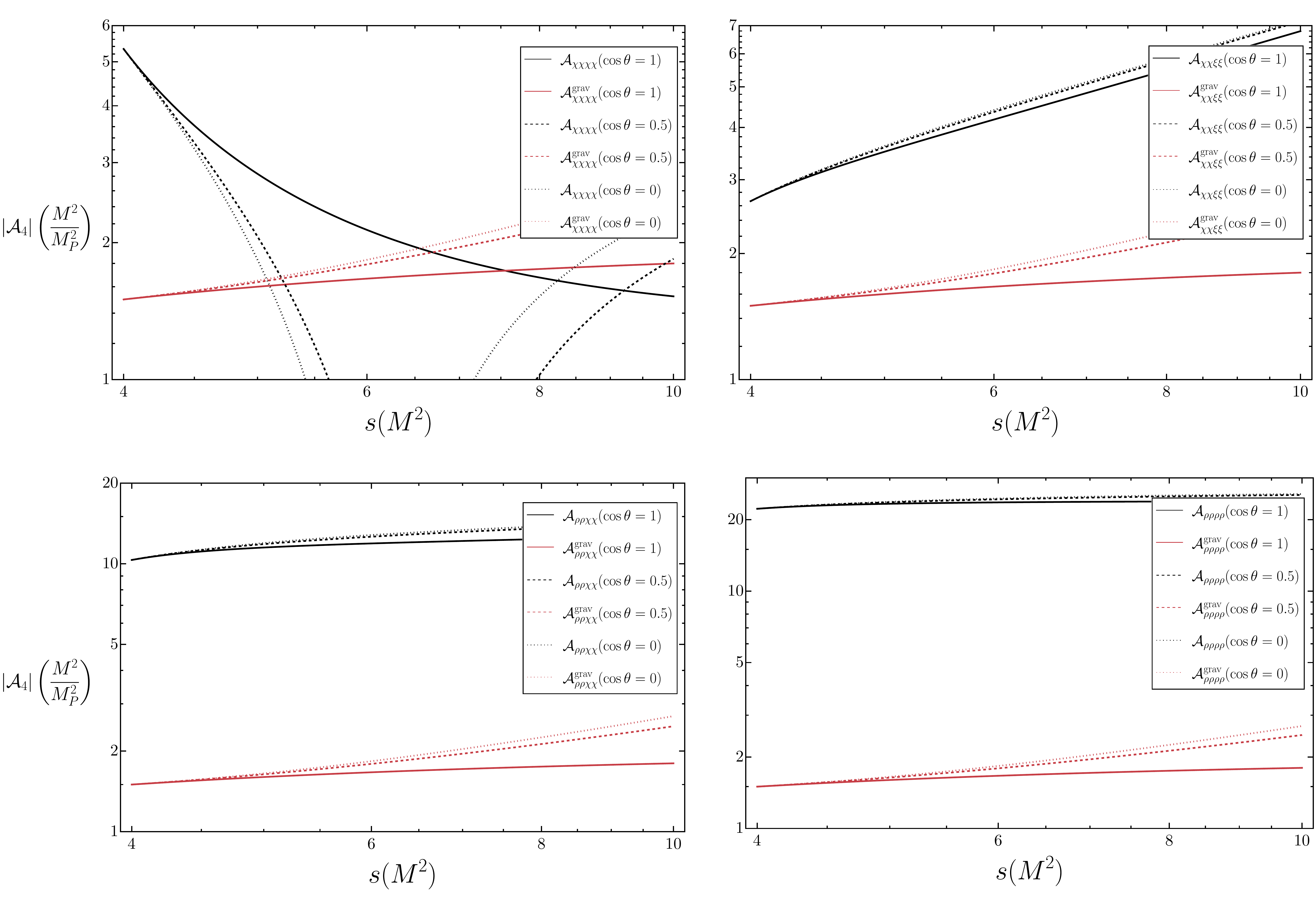}
    \caption{As in Fig.~\ref{ampi1} for the Cecotti inflationary model. }
    \label{ampi2}
\end{figure}

\begin{figure}[ht!]
    \centering   \includegraphics[width=0.9\columnwidth]{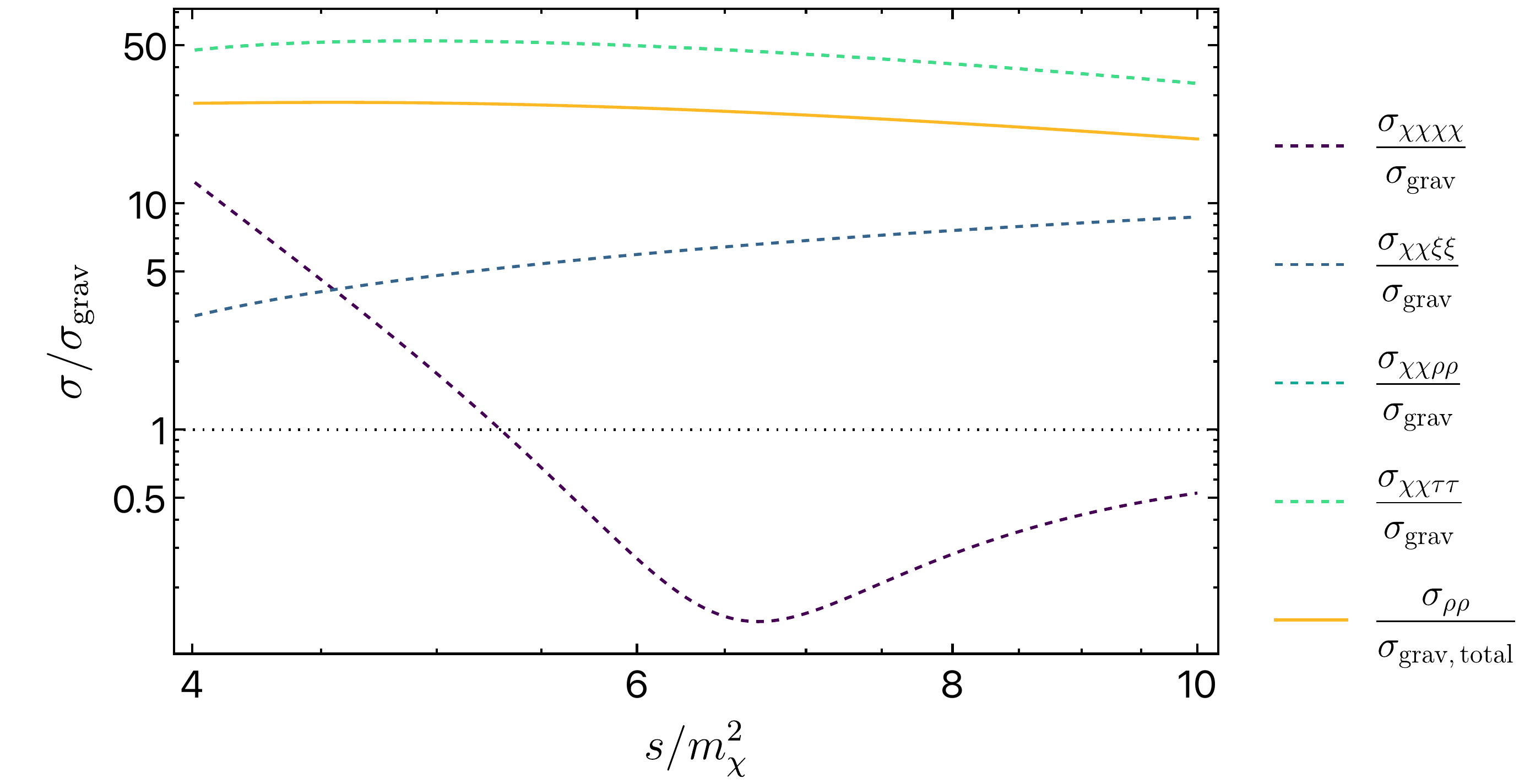}
\includegraphics[width=0.9\columnwidth]{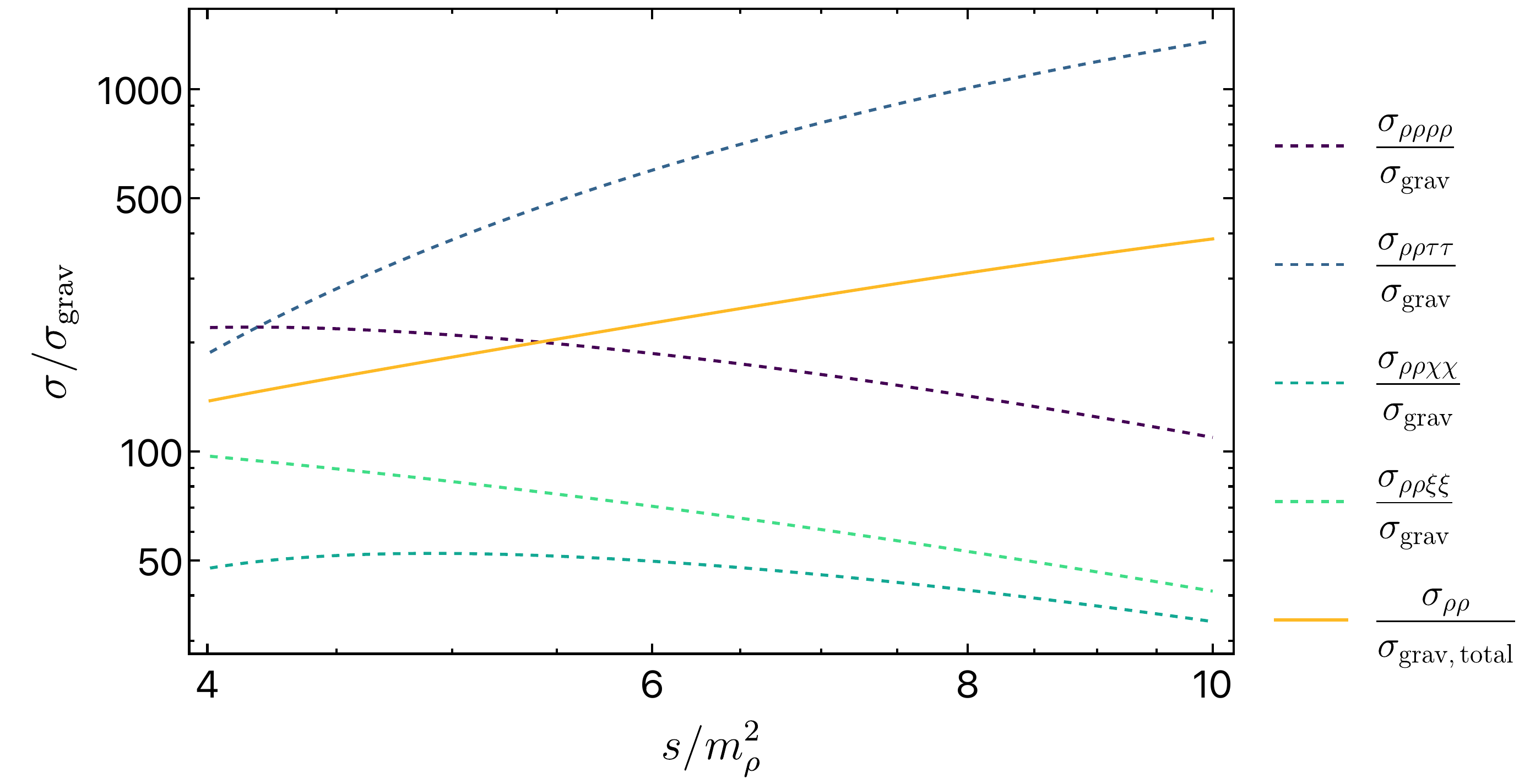}
    \caption{As in Fig.~\ref{csinf1} for the Cecotti inflationary model. }
    \label{csinfl2}
\end{figure}

At the potential minimum, the masses of all four real scalar fields are equal to $M$.
The amplitudes and cross sections for $\chi \chi$ and $\rho \rho$ scattering at the minimum are shown in Figs.~\ref{ampi2} and \ref{csinfl2}, respectively with $\Lambda_\phi = M_P$. In this case all four final states are available in $\chi \chi$ scattering and both cross sections are larger than the corresponding gravitational cross section.

\begin{figure}[ht!]
    \centering   \includegraphics[width=0.9\columnwidth]{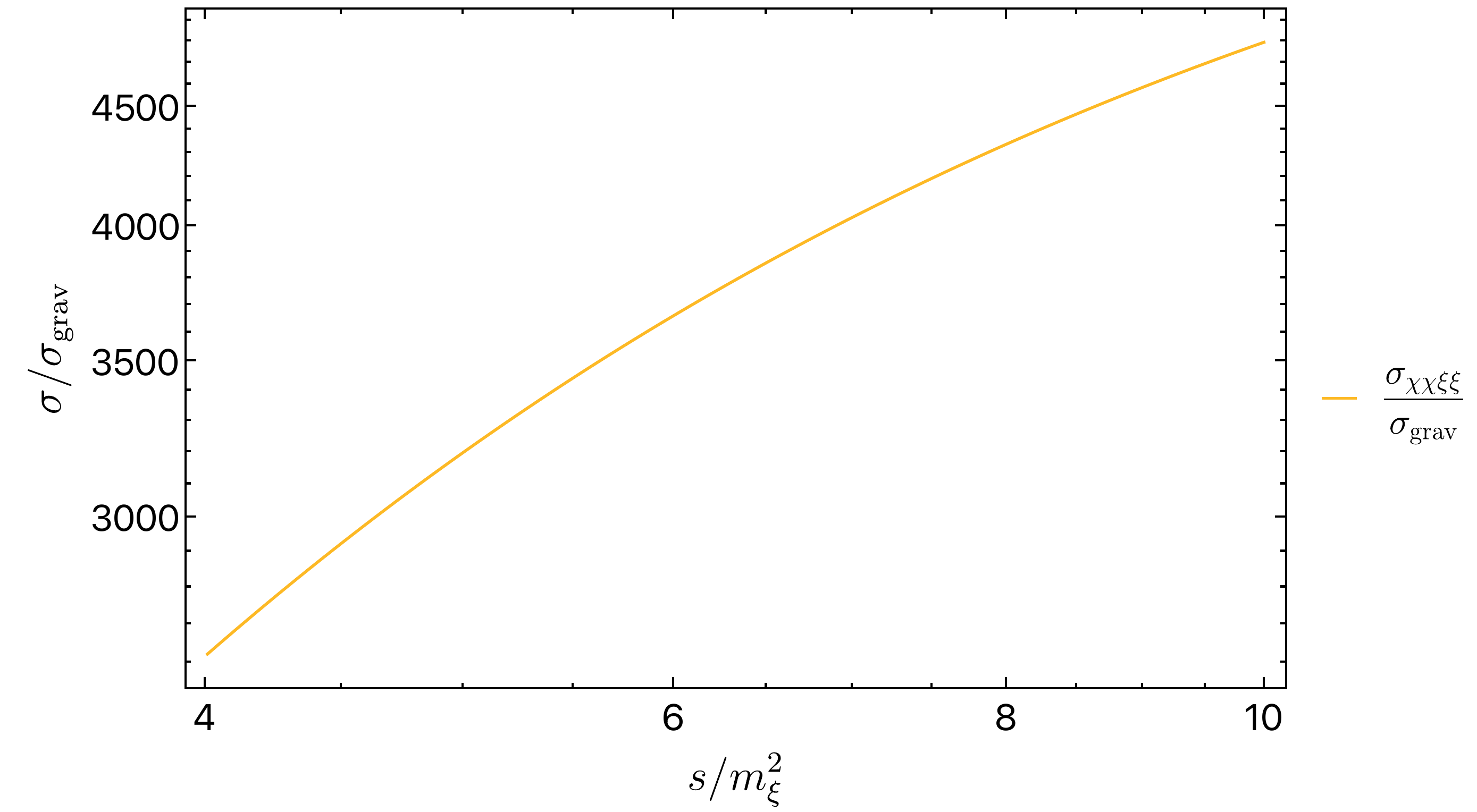}
    \caption{As in Fig.~\ref{csinfl2} for scattering when the background field values take the limit, $v \to \infty$ and $p = 0$. }
    \label{cstab}
\end{figure}

The saddle point in the Cecotti model occurs when $\phi = 0$ and $T \to \infty$ or when $p = 0$ and $v \to \infty$. In this limit the masses of $\rho$ and $\tau$ both diverge and the light fields have masses, $m_\chi = 0$ and $m_\xi = M$.  The cross section for $\chi \chi \to \xi \xi$ and the equal reverse process  at the saddle point is shown in Fig.~\ref{cstab}. In contrast to the previous inflationary example, the cross section in this model always satisfies the $\sigma_{ai}$SWGC. 

Note that in this model during the inflationary period, the kinetic terms approach a zero, instead of a pole seen in the previous Wess-Zumino inflationary model. It was argued in \cite{Scalisi:2018eaz}
that such models respect the SDC~\cite{Ooguri:2006in} and that they have better prospects for string realizations. Other swampland conjectures related to no-scale supergravity were discussed in \cite{Rasulian:2021wny}.

\section{Conclusions}
\label{sec:concl}

When low-energy effective field theories are embedded into a theory of quantum gravity, nontrivial constraints can be obtained on the  available parameter space. Based on black hole physics, the WGC provides one of the most surprising constraints on a U(1) gauge theory when coupled to gravity by requiring the existence of a WGC charged particle with $m\leq g q M_P$. Interestingly, this constraint can also be understood as the condition that the Coulomb interaction is stronger than gravity. The problems that arise when the WGC is violated are evident.  This conjecture can be modified to include scalar fields that couple to the WGC particle in the U(1) gauge theory. However, when this conjecture is generalized to only include scalar fields (SWGC), arguments based on black hole physics no longer apply.
 Indeed, the absence of a charge would violate the conjecture unless repulsive interactions are introduced as in Eq.~(\ref{wgcr}). In contrast, in the SWGC, the attractive Coulomb-like forces are moved to the opposite side of the inequality as in Eq.~(\ref{swgc}), but we lose the obvious problem which arises when the conjecture is violated.
Instead, as advocated in Ref.~\cite{Gonzalo:2019gjp}, one can adopt the principle that gravity should always be the weakest force (sSWGC). 

One way to test the sSWGC at arbitrary field values, involves comparing the four-point scattering amplitudes from scalar interactions with those arising from gravity (specifically at a particular energy scale). For the case of multiple scalar fields with noncanonical kinetic terms, which are generic in no-scale supergravity models, we employ the methods of Ref.~\cite{dgov} to canonically normalize the scalar fields. This is done by transforming the scalar fields to the KNC (or RNC) basis, greatly simplifying the calculation of scattering amplitudes of the asymptotic states. Since the scattering is computed at arbitrary field values, all terms in the scalar potential $V(\phi)$ should be included in the calculation of the amplitudes. In particular, linear terms are generated at generic field values away from the potential extrema. However, as shown in Ref.~\cite{dgov}, the scattering amplitudes do not remain invariant under a field transformation, if the linear terms are neglected. Instead, as expected, the amplitudes are invariant at the potential extrema (where the linear terms vanish). To the extent that a scattering process can be defined at non-extremal field values (by neglecting the time-dependence of the background value) we also provide evidence that summing all tadpole contributions reproduces the scattering that would be obtained at the potential minima. This suggests that scattering amplitudes only need to be compared at the extrema of the scalar potential.

The fact that scattering amplitudes are covariant, depending on the energy scale and angular direction, further suggests one should integrate over the scattering angle in order to compare the scalar interactions with gravity ($\sigma$SWGC). We consider several examples of real (and complex) scalars coupled to gravity and sum over the possible final states. This effectively yields a scattering cross section that can then be compared to gravity, providing a much more general condition compared to the usual comparison with nonrelativistic forces. This also mimics the reformulation of the sSWGC in Ref~\cite{Gonzalo:2020kke}, which argued that the gravitational interaction rates should always be smaller than the pair production at threshold. Therefore, adopting the condition that $\sigma_{\rm scalar} > \sigma_{\rm gravity}$ to test the sSWGC, we find no violations in our toy examples depending on the specific formulation of the conjecture.

We also applied our methods to two realistic inflationary models motivated from no-scale supergravity. At the potential minimum, 
both models satisfied the $\sigma_{ai}$SWGC. Namely for each of the four real fields in the model, the inclusive scalar cross section is larger than the inclusive gravitational cross section.
Along one real direction, both models of inflation are Starobinsky-like.  Therefore, each has a saddle point extremum in the limit that the canonical inflaton tends to infinity.
At this extremal limit, we found
a violation of our $\sigma_{ai}$SWGC and $\sigma_{a1}$SWGC conditions in the Wess-Zumino-like inflation example. Namely, the cross section for inflaton scattering (the inflaton is massless in this limit)
is smaller than the gravitational cross section for a certain range of center of mass energies. Indeed, both conjectures are satisfied in both the high and low energy limits. 
On the other hand, the $\sigma_{1i}$SWGC is satisfied as it requires only one initial state inclusive cross section to be larger than the gravitational cross section. 
Thus, the model may not be relegated to the swampland. In addition, we reiterate that we have not accounted for the fact that the background metric is de Sitter in this limit, whereas we have assumed a Minkowski background throughout the paper.

Previous attempts at testing scalar field theories using some form of the SWGC, were based on amplitudes or properties of the scalar potential at all points in field space. In the latter case, the presence of tadpole terms (necessarily present away from extrema - by definition) were neglected. In \cite{dgov}, we provided a procedure for calculating amplitudes for non-trivial field space manifold utilizing either Riemann or K\"ahler normal coordinates. The latter is particularly well suited for calculations in the context of non-minimal supergravity models such as no-scale supergravity. We also showed that amplitudes calculated when neglecting tadpole contributions were not field redefinition invariant. Using a simple example, this point was  clarified in section \ref{sec:simptad}, while in section \ref{sec:perttad} we showed that for small deviations from an extremum,
the inclusion of tadpole contributions leads to amplitudes which are given by their extremal expressions. We also argued more generally that from the Newton tangent method, amplitudes at arbitrary field values should be equivalent to the corresponding amplitudes evaluated at the extrema of the potential. 

Nevertheless, we are left with a number of questions. Which (if any) of the several versions of the $\sigma$SWGC is a true test of whether a scalar field theory is in the swampland? We have seen in nearly every example, that a comparison of the scalar and gravitational amplitudes is not a suitable test because at certain center of mass energies and scattering angles, the scalar amplitude changes sign and therefore vanishes at some energies. Perhaps this is avoided in the low energy limit, but it is not clear why this limit would be appropriate as we do not rely on the existence of bound states (as in the WGC). Even when comparing amplitudes, it is unclear if this should be applied for
each distinct initial and final state, or if to satisfy the conjecture, it is sufficient to have 
a large (low energy) amplitude for a single final state for any initial state, or if it is necessary to satisfy the inequality for all initial states and at least one  (or all) final state(s). 

The problem of scattering angle is removed by calculating and comparing cross sections. We have outlined several possible forms for the conjecture expressed in terms of scattering cross sections. 
The strongest constraint 
would require satisfying the inequality for all initial and final states. However, with the exception of the KKLT model discussed in section~\ref{sec:KKLT}, this version of the conjecture is violated in every model considered, as there is always a specific initial and final state where the cross section is smaller than the gravitational cross section. 
Finally, it is also not clear which states are necessary to consider.
All of the toy models (including the inflationary models) consist of one or two chiral fields. However, in addition, there are all of the Standard Model fields and in a supersymmetric theory would include at least 98 Standard Model (real) scalars. Should these be included in a gravitational inclusive cross-section? Should the conjecture be applied only in the low energy limit, or at all scattering energies? In the context of supergravity, which interactions should be considered gravitational? It is clear that a better understanding of the nature of the conjecture as applied to scalar fields is needed if we wish to discriminate between viable UV theories and the swampland. 

\section*{Acknowledgements}
We would like to thank E. Gonzalo, L. Ibanez, M. Luty, and E. Palti for useful discussions. The work of T.G and K.A.O.~was supported in part by DOE grant DE-SC0011842 at the University of Minnesota. The work of T.G. was performed in part at the Aspen Center for Physics, which is supported by National Science Foundation grant PHY-1607611. The work of S.V. was supported in part by DOE grant DE-SC0022148.

\end{document}